\documentclass[10pt,a4paper,oneside]{article}

\usepackage[T1]{fontenc}
\usepackage[latin1]{inputenc}
\usepackage[english]{babel}
\usepackage{fancyhdr}
\usepackage{amsmath}
\usepackage{amsfonts}
\usepackage{amssymb}
\usepackage{amscd}
\usepackage{amsbsy}
\usepackage{lastpage}
\usepackage{enumitem}
\usepackage{mathrsfs}% script fonts
\usepackage[sort&compress,longnamesfirst]{natbib}
\usepackage{color}
\usepackage[top=4cm,bottom=3cm,left=4cm,right=3cm]{geometry}
\usepackage[bookmarks=false,colorlinks=true,linkcolor=UniBlue,citecolor=UniRed,filecolor=UniBlue,urlcolor=UniBlue,pdftitle={Non-concave utility maximisation on the non-negative real axis in discrete time},pdfauthor={Laurence Carassus, Mikl\'{o}s R\'{a}sonyi, and Andrea Meireles Rodrigues},%
pdfsubject={This work treats the discrete-time asset allocation problem, in an arbitrage-free but generically incomplete financial model, when the investor has a possibly non-concave utility function and wealth is restricted to remain non-negative. Under straightforward conditions, we use a dynamic programming framework to establish the existence of an optimal portfolio.},%HEREEE
pdfkeywords={Attainability;\ } {Discrete-time\ models;\ } {Dynamic\ programming;\ } {Finite\ horizon;\ } {Incomplete\ markets;\ } {Non-concave\ utility;\ } {Optimal\ portfolio}]{hyperref}%HEREEE
\usepackage{amsthm}
\usepackage[backgroundcolor=white,linecolor=UniRed,bordercolor=UniRed,textsize=footnotesize,textwidth=2cm%
,disable% activate disable for the FINAL version
]{todonotes}% \todo{}, to include margin comments in .pdf
\definecolor{UniBlue}{RGB}{0,50,95}
\definecolor{UniRed}{RGB}{193,0,67}

\bibpunct{[}{]}{;}{n}{,}{,}

\reversemarginpar% for todo notes on the left margin

\newcommand{\bbOne}{1\hspace*{-0.8ex}1}

\newcommand{\rcd}[3][]{\mathbb{P}^{\left. #2 \right|\mathscr{#3}_{#1}}}

\newcommand{\Proj}[2][\Omega]{\operatorname{Proj}_{#1}\!\left(#2\right)}

\DeclareMathOperator{\supp}{supp}
\DeclareMathOperator{\aff}{aff}

\DeclareMathOperator*{\esssup}{ess\,sup}
\DeclareMathOperator{\gph}{gph}

\numberwithin{equation}{section} %section

\newtheorem{theorem}{Theorem}[section]%section
\newtheorem{proposition}[theorem]{Proposition}
\newtheorem{lemma}[theorem]{Lemma}

\newtheorem{assumption}[theorem]{Assumption}

\theoremstyle{definition}
\newtheorem{definition}[theorem]{Definition}

\theoremstyle{remark}
\newtheorem{remark}[theorem]{Remark}

\newtheoremstyle{example}% name of the style to be used
  {0.5\topsep}% measure of space to leave above the theorem. E.g.: 3pt
  {0.5\topsep}% measure of space to leave below the theorem. E.g.: 3pt
  {\normalfont}%{\itshape}% name of font to use in the body of the theorem
  {0pt}% measure of space to indent
  {\itshape}%{\normalfont}% name of head font
  {.}% punctuation between head and body
  {5pt plus 1pt minus 1pt}% space after theorem head; " " = normal interword space
  {}% Manually specify head
\theoremstyle{example}
\newtheorem{example}[theorem]{Example}

\def\shorttitle{Non-concave utility maximisation on the positive real axis in discrete time}

\title{Non-concave utility maximisation\\on the positive real axis in discrete time%
\thanks{%The authors wish to thank an anonymous referee for a careful reading of their manuscript and valuable suggestions. %
L.~Carassus thanks LPMA (UMR7599) for support.
A.\,M.~Rodrigues gratefully acknowledges the financial support of FCT-Funda\c{c}\~{a}o para a Ci\^{e}ncia e Tecnologia (Portuguese Foundation for Science and Technology) through the Doctoral Grant SFRH/BD/69360/2010. Part of this research was carried out while M.~R\'{a}sonyi.\ and A.\,M.~Rodrigues were affiliated with the School of Mathematics, University of Edinburgh, Scotland, U.K.}}

\date{\today}

\author{Laurence Carassus\footnote{LMR (EA 4535, CNRS FR 3399ARC), Universit\'e Reims Champagne-Ardenne, Moulin de la Housse -- BP1039, 51687 Reims cedex 2, France. \mbox{E-mail}:~\texttt{\href{mailto:laurence.carassus@univ-reims.fr}{laurence.carassus@univ-reims.fr}}}
\and Miklós~Rásonyi\footnote{MTA Alfr\'{e}d R\'{e}nyi Institute of Mathematics, Budapest,
Hungary. The second author is also affiliated with P\'azm\'any P\'eter Catholic University, Budapest, Hungary. \mbox{E-mail}:~\texttt{\href{mailto:rasonyi@renyi.mta.hu}{rasonyi@renyi.mta.hu}}}
\and Andrea~M.~Rodrigues\footnote{Dublin City University, Dublin, Ireland. \mbox{E-mail}:~\texttt{\href{mailto:Andrea.MeirelesRodrigues@dcu.ie}{Andrea.MeirelesRodrigues@dcu.ie}}}}

\begin{document}

\fancypagestyle{plain}{ %
  \fancyhf{} % remove everything
  \renewcommand{\headrulewidth}{0pt} % remove lines as well
  \renewcommand{\footrulewidth}{0pt}
  %\cfoot{Page \thepage/\pageref{LastPage}}
}
\thispagestyle{plain}

\maketitle

\begin{abstract}	
	We treat a discrete-time asset allocation problem in an arbitrage-free, generically
	incomplete financial market, where the investor has a possibly non-concave utility function and wealth is
	restricted to remain non-negative. Under easily verifiable conditions, we
	establish the existence of optimal portfolios.
	
\end{abstract}

%%%%%%%%%%%%%%%%%%%%%%%%%%%%%%%%%%%%%%%%%%%%%%%%%%%%%%%%%%%%%%%%%%%%%%%%%%%%%%%%%%%%%%%%%%%%%%%%%%%%%%%%%%%%%%%%%%%%%%%%%%%%%%%%%%%%%%%%%%
%%%%%%%%%%%%%%%%%%%%%%%%%%%%%%%%%%%%%%%%%%%%%%%%%%%%%%%%%%%%%%%%%%%%%%%%%%%%%%%%%%%%%%%%%%%%%%%%%%%%%%%%%%%%%%%%%%%%%%%%%%%%%%%%%%%%%%%%%%
%%                                                                                                                                      %%
%% Main Body                                                                                                                            %%
%%                                                                                                                                      %%
%%%%%%%%%%%%%%%%%%%%%%%%%%%%%%%%%%%%%%%%%%%%%%%%%%%%%%%%%%%%%%%%%%%%%%%%%%%%%%%%%%%%%%%%%%%%%%%%%%%%%%%%%%%%%%%%%%%%%%%%%%%%%%%%%%%%%%%%%%
%%%%%%%%%%%%%%%%%%%%%%%%%%%%%%%%%%%%%%%%%%%%%%%%%%%%%%%%%%%%%%%%%%%%%%%%%%%%%%%%%%%%%%%%%%%%%%%%%%%%%%%%%%%%%%%%%%%%%%%%%%%%%%%%%%%%%%%%%%

\pagestyle{fancy}
\fancyhf{} % remove everything
\renewcommand{\headrulewidth}{0pt} % remove lines as well
\renewcommand{\footrulewidth}{0pt}
\cfoot{Page \thepage/\pageref{LastPage}}
\chead{{\small \shorttitle}}

%%%%%%%%%%%%%%%%%%%%%%%%%%%%%%%%%%%%%%%%%%%%%%%%%%%%%%%%%%%%%%%%%%%%%%%%%%%%%%%%%%%%%%%%%%%%%%%%%%%%%%%%%%%%%%%%%%%%%%%%%%%%%%%%%%%%%%%%%%
%%                                                                                                                                      %%
%% Introduction and summary                                                                                                             %%
%%                                                                                                                                      %%
%%%%%%%%%%%%%%%%%%%%%%%%%%%%%%%%%%%%%%%%%%%%%%%%%%%%%%%%%%%%%%%%%%%%%%%%%%%%%%%%%%%%%%%%%%%%%%%%%%%%%%%%%%%%%%%%%%%%%%%%%%%%%%%%%%%%%%%%%%

\section{Introduction}\label{sec:Intro}

We consider investors trading in a multi-asset and discrete-time financial market who are aiming to
maximise their expected utility from terminal wealth. If the utility function $u$ is defined on the non-negative
half-line, is concave, and the problem has a finite value function, then there is always such a strategy,
see \citet{rsch06}. In a general semimartingale model  one needs to assume, in addition, that the so-called ``asymptotic elasticity
at $+\infty$'', denoted by $AE_+(u)$, is less than one in order to obtain an optimal portfolio for the utility maximization problem,
%: one need to assume that , $AE_+<1$,
see \citet{kramkov99} and Remark \ref{obs:growthu} below. In the utility maximisation context, conditions on the asymptotic elasticity (which were
first used in
\citet{ck,karatzas_et_al,kramkov99}) have become standard in the literature.

In this paper we want to remove the assumptions of concavity and smoothness that are usually made on $u$.
Why? Several reasons can be invoked. The first one is quite clear: the investor can change her perception of
risk above a certain level of wealth. One can also consider the problem of  optimizing performance
at some level $B$: one penalizes loss under $B$ with a loss function and one maximizes gain after $B$ with a
gain function. These illustrations are typical examples of a piecewise concave function. This kind of
problem has been addressed in the
complete case by \citet{carassus-pham} and in a pseudo-complete market by \citet{reichlin2011}.

Other examples of non-concave utility functions are the so-called ``\emph{S}-shaped'' functions. These
appeared in cumulative prospect theory (or CPT for short, see \citet{kahneman79,tversky92}). This theory asserts that the problem's mental
representation is important: agents analyse their gains or losses
with respect to a given stochastic reference point $B$ rather than
with respect to zero and they take  potential losses more
into account than potential gains. Note that in this paper, in contrast to cumulative prospect theory,
we do not allow investors to distort the
probability measure by a transformation function of the cumulative
distributions. The case of ``\emph{S}-shaped'' functions was studied by \citet{berkelaar04} in a complete market
setting.

In the present paper we provide mild sufficient conditions (involving asymptotic elasticity) on a possibly non-concave, non-differentiable and random utility function defined on the non-negative half-line which guarantee the existence of an optimal strategy.
By treating multi-step discrete-time markets, we cover a
substantial class of incomplete models which can be fitted to arbitrary econometric data.

The case of (non-random) utilities defined on the whole real line was treated in \citet{carassusrasonyi2012}, but so far there were no general results in the case of the non-negative half-line; in the present setting we are only aware of Chapter~IV of \citet{reichlinthesis},
where existence results were proved for some very specific market models.

We finish by listing some references in the case of probability distortions in the spirit of
cumulative prospect theory, because those results can be applied to our setup with weight functions equal to
the identity. In the incomplete discrete-time setting, the papers of
\citet{carassusrasonyi2011,rasonyirodriguez2013} study quite specific utility functions.
In continuous-time studies, all the references make the assumption that the market is complete:
see \citet{jin2008}, \citet{carlier-dana2011}, \citet{cd11}, \citet{rasonyi2013}.

%\footnote{Actually, the assumptions of \citet{rasonyi2013} require slightly less than completeness, but this does not lead to a significant improvement in generality.}.

A brief outline of this article is as follows.
Section~\ref{sec:SetUp} is dedicated to specifying the market model and to introducing the relevant notations.
In Section~\ref{ExistTheo} we formulate our main result. Next, in Section~\ref{sec:OneStep} we examine the problem in a one-step setting, whilst in Section~\ref{sec:Dynamic} we prove our main result, using a dynamic programming approach.
For the sake of a simple presentation, the proofs of some auxiliary results are collected in Section~\ref{sec:Aux}.

%%%%%%%%%%%%%%%%%%%%%%%%%%%%%%%%%%%%%%%%%%%%%%%%%%%%%%%%%%%%%%%%%%%%%%%%%%%%%%%%%%%%%%%%%%%%%%%%%%%%%%%%%%%%%%%%%%%%%%%%%%%%%%%%%%%%%%%%%%
%%                                                                                                                                      %%
%% Notation and set-up                                                                                                                  %%
%%                                                                                                                                      %%
%%%%%%%%%%%%%%%%%%%%%%%%%%%%%%%%%%%%%%%%%%%%%%%%%%%%%%%%%%%%%%%%%%%%%%%%%%%%%%%%%%%%%%%%%%%%%%%%%%%%%%%%%%%%%%%%%%%%%%%%%%%%%%%%%%%%%%%%%%

\section{Notation and set-up}\label{sec:SetUp}

%=========================================================================================================================================
%
% The market
%
%=========================================================================================================================================

\subsection{The market}

In what follows, we shall consider a frictionless and totally liquid financial market model with finite trading horizon $T\in\mathbb{N}$, in which the current time is denoted by $0$ and trading is assumed to occur only at the dates $\left\{0,1,\ldots,T\right\}$.

As usual, the uncertainty in the economy is characterised by a complete probability space	
$\left( \Omega,\mathscr{F},\mathbb{P} \right)$, where $\mathscr{F}$ is a $\sigma$-algebra on
the sample space $\Omega$, and $\mathbb{P}$ is the underlying probability measure
(to be interpreted as the physical probability). Moreover, all
the information accruing to the agents in the economy is described by a discrete-time
filtration $\mathbb{F}=\left\{\mathscr{F}_{t}\text{; }t\in\left\{0,1,\ldots,T\right\}\right\}$
such that $\mathscr{F}_{0}$ coincides with the family of all $\mathbb{P}$-null sets.
Finally, we assume also that $\mathscr{F}=\mathscr{F}_{T}$.

Next, we fix an integer $d>0$ and consider a
process $S=\left\{S_{t}\text{; }t\in\left\{0,1,\ldots,T\right\}\right\}$, so that $S_{t}$ represents the time-$t$
prices of $d$ traded risky assets. Denoting by $\Xi^{n}_{t}$ the family of all $\mathscr{F}_{t}$-measurable
random vectors $\xi:~\Omega\rightarrow\mathbb{R}^{n}$ for each $n\in\mathbb{N}$ and each
$t\in\left\{0,1,\ldots,T\right\}$, we assume that $S_{t}\in\Xi^{d}_{t}$ for every
$t\in\left\{0,1,\ldots,T\right\}$, i.e., $S$ is $\mathbb{F}$-adapted.
We shall also assume, without loss of generality, that the risk-free asset in this economy has constant
price equal to one at all times. Finally, for each $t\in\left\{1,\ldots,T\right\}$, we define $\Delta S_{t}\triangleq S_{t}-S_{t-1}$.

We recall that a \emph{self-financing portfolio} is a process $\phi=\left\{\phi_{t}\text{; }t\in\left\{1,\ldots,T\right\}\right\}$, %\footnote{Here, the superscript $\top$ denotes matrix transposition.}, %
with $\phi_{t}\in\Xi^{d}_{t-1}$ for all $t\in\left\{1,\ldots,T\right\}$, and its \emph{wealth process} $\Pi^{\phi}=\left\{\Pi^{\phi}_{t}\text{; }t\in\left\{0,1,\ldots,T\right\}\right\}$ satisfies, for every $t\in\left\{1,\ldots,T\right\}$,% $\Pi_{t}^{\phi}=\Pi_{0}^{\phi}+\sum_{s=1}^{t}\left\langle \phi_{s},\Delta S_{s} \right\rangle$ a.s.\footnote{We recall that $\left\langle \cdot,\cdot \right\rangle$ and $\left\Vert\cdot\right\Vert$ respectively denote the \emph{Euclidean inner product} and the \emph{Euclidean norm}.} %
\begin{equation*}
	\Pi_{t}^{\phi}=\Pi_{0}^{\phi}+\sum_{s=1}^{t}\left\langle \phi_{s},\Delta S_{s} \right\rangle\text{ a.s.}
\end{equation*}
Here $\langle\cdot,\cdot\rangle$ denotes scalar product in $\mathbb{R}^d$ and $\Vert\cdot\Vert$
is the corresponding Euclidean norm.
We denote by $\Phi$ the class of all self-financing portfolios.
In addition, we shall impose the following trading constraint: the value of a portfolio should not be allowed to become
strictly negative. So we say that a portfolio $\phi\in\Phi$ is \emph{admissible} for $x_{0}\geq 0$ (and we write $\phi\in\Psi\!\left(x_{0}\right)$) if, for every $t\in\left\{1,\ldots,T\right\}$, the inequality $\Pi^{\phi}_{t}\geq 0$ holds a.s.
with $\Pi_0^{\phi}=x_0$. Because of budget constraints, such a restriction is natural and frequently imposed, see e.g.\ \citet{kramkov99,rsch06}.
%\begin{remark}
%	Neither \citet{rs05} nor \citet{carassusrasonyi2012} place any admissibility restriction on the portfolios, and therefore they allow portfolios whose value processes not only may take strictly negative values with positive probability, but also are not necessarily bounded from below. We observe, however, that our admissibility condition, despite considerably restricting the set of admissible portfolios, % (cf.\ Remark~2.1 in \citet{carassusrasonyi2012}), %
%	is frequently adopted in the literature, as it has
%the natural interpretation of forbidding strictly negative positions (i.e., debts).
%\end{remark}

The following no arbitrage assumption stipulates that no investor should be allowed to make a profit out of nothing and without risk, even with a budget constraint.
\begin{assumption}\label{as:NA}
	The market does not admit arbitrage, i.e.,
	\begin{equation*}\tag{NA}
		\text{for all }x_{0}\geq0 \text{, if }\phi\in\Psi\!\left(x_{0}\right)\text{ with }\Pi_{T}^{\phi}\geq x_{0}\text{ a.s., then }\Pi_{T}^{\phi}=x_{0}\text{ a.s.}
		\label{eq:NA}
	\end{equation*}
\end{assumption}
\begin{remark}
\label{rmqnaplus}
It is proved in Proposition~1.1 of \citet{rsch06} that (NA) is equivalent to the classical no arbitrage condition: $\forall \phi \in \Phi$,
$\Pi_{T}^{0,\phi}\geq 0$  a.s. implies that $\Pi_{T}^{0,\phi}=0$  a.s. where  $\Pi_{T}^{0,\phi}$ stands for the wealth process associated to $\phi$ when starting with a zero initial wealth i.e. $\Pi_{0}^{\phi}=0$.
\end{remark}

Now fix $t\in\left\{1,\ldots,T\right\}$. We know that there exists a \emph{regular conditional distribution} of $\Delta S_{t}$ with respect to $\mathscr{F}_{t-1}$ under the physical measure $\mathbb{P}$, %(see \Cref{prop:existrcd}), %
%(see e.g.\ Theorem~10.2.2 in \citet[pp.~345--346]{dudley2004}), %
which we shall denote by $\rcd[t-1]{\Delta S_{t}}{F}$. By modifying on a $\mathbb{P}$-null set,
we may and will assume that $\rcd[t-1]{\Delta S_{t}}{F}(\cdot,\omega)$ is a probability
for \emph{all} $\omega\in\Omega$.
Let $D_{t}\!\left(\omega\right)$ denote the affine hull in $\mathbb{R}^{d}$ of the support of
$\rcd[t-1]{\Delta S_{t}}{F}(\cdot,\omega)$. %, otherwise simply set $D_{t}\!\left(\omega\right)\triangleq \mathbb{R}^{d}$. Clearly, %each $D_{t}\!\left(\omega\right)$ is an affine subspace of $\mathbb{R}^{d}$. We further observe that, %
%for every $\omega\notin\overline{\Omega}_{t}$, $D_{t}\!\left(\omega\right)$ is actually a linear space. %
%so too is $D_{t}\!\left(\omega\right)$ %
It follows from Theorem~3 in \citet{jacod98} that, under (NA), $D_{t}\!\left(\omega\right)$ is actually a linear space for $\mathbb{P}$-almost every $\omega$.
%($\mathbb{P}$-a.e.) $\omega\in\overline{\Omega}_{t}$.

%\begin{proposition}\label[prop]{prop:linsub}
%	Suppose \eqref{eq:NA} is verified. Then, for every $t\in\left\{1,\ldots,T\right\}$, there exists a subset $\widehat{\Omega}_{t}$ of $\overline{\Omega}_{t}$ satisfying all of the following conditions,
%	\begin{enumerate}[label=\emph{(\roman*)}]
%		\item
%		$\widehat{\Omega}_{t}$ belongs to the $\sigma$-algebra $\mathscr{F}_{t-1}$,
%		
%		\item
%		$\mathbb{P}\!\left(\overline{\Omega}_{t}\setminus\widehat{\Omega}_{t}\right)=0$,
%		
%		\item
%		for every $\omega\in\widehat{\Omega}_{t}$, the affine space $D_{t}\!\left(\omega\right)$ is actually a linear subspace of $\mathbb{R}^{d}$.
%	\end{enumerate}
%\end{proposition}

%\begin{proof}
%	See{\todo{HEREEE}}.
%\end{proof}

%Furthermore, for each fixed $t\in\left\{1,\ldots,T\right\}$, we define an important family of functions.
Given any $\mathscr{F}_{t-1}$-measurable random variable %$H$ verifying %
$H\geq 0$ a.s.\ (which can also be some constant $x \geq 0$), we set
\begin{equation*}
	\Xi^{d}_{t-1}\!\left(H\right)\triangleq \left\{\xi\in\Xi^{d}_{t-1}\text{: }H+\left\langle \xi,\Delta S_{t} \right\rangle\geq 0\text{ a.s.}\right\}.
\end{equation*}

%In the particular case where $H=x$ a.s.\ for some $x\geq 0$, we have
%\begin{equation*}
%	\Xi^{d}_{t-1}\!\left(x\right)\triangleq \left\{\xi\in\Xi^{d}_{t-1}\text{: }x+\left\langle \xi,\Delta S_{t} \right\rangle\geq 0\text{ %a.s.}\right\}.
%\end{equation*}
We take $\widetilde{\Xi}^{d}_{t-1}$ to be the class of all random vectors $\xi\in\Xi^{d}_{t-1}$ such that $\xi\!\left(\omega\right)\in D_{t}\!\left(\omega\right)$ for $\mathbb{P}$-a.e.\ $\omega$.
The notation $\tilde{\Xi}_{t-1}^d(H)$ is self-explanatory.
%, and by abuse of language we shall write that `$\xi$ belongs to $D_{t}$ a.s.'.

%We end this subsection with an alternative and rather useful characterisation of the no-arbitrage condition \eqref{eq:NA}.
\begin{proposition}\label{prop:quantNA}
	The following two statements are equivalent,
	\begin{enumerate}[label=\emph{(\roman*)}]%,leftmargin=\parindent+2.5mm]
		\item
		\eqref{eq:NA} holds true;
		
		\item
		for every $t\in\left\{1,\ldots,T\right\}$, there exist $\mathscr{F}_{t-1}$-measurable
		random variables  $\beta_{t}>0, \kappa_{t}>0$ a.s. such that,
		for every $\xi\in\widetilde{\Xi}^{d}_{t-1}$, the inequality
	\begin{equation}\label{eq:NAbk}
		\mathbb{P}\!\left(\left.\left\langle \xi,\Delta S_{t} \right\rangle\leq-\beta_{t}\left\Vert \xi\right\Vert\right|\mathscr{F}_{t-1}\right)\geq\kappa_{t}% \text{ a.s.\ on }\widetilde{\Omega}_{t}\triangleq\left\{\omega\in\widehat{\Omega}_{t}\text{: }D_{t}\!\left(\omega\right)\neq\left\{\vect{0}\right\}\right\}
	\end{equation}%
	%for all $\xi\in\widetilde{\Xi}^{d}_{t-1}$.
	holds a.s.
	\end{enumerate}
\end{proposition}

\begin{proof}
	This follows from Proposition 3.3 in \citet{rs05}  and Remark \ref{rmqnaplus} above (see also Proposition~1.1 in \citet{rsch06}).%, combined with Jacod and Shiryaev.
\end{proof}

\begin{remark}
	%\begin{enumerate}[label=\emph{(\roman*)}]
		%\item
		%It follows from the proof of \Cref{prop:quantNA} that, for every $t\in\left\{1,\ldots,T\right\}$, we have $\beta_{t}\leq 1$ a.s.\ as well. It is also trivial that, for every $t\in\left\{1,\ldots,T\right\}$, the inequality $\kappa_{t}\leq 1$ holds a.s.
		
		%\item
		%An inspection of the proof of \Cref{prop:quantNA} allows us to conclude that, in the particular case where $S$ has independent increments, the random variables $\beta_{t}$ and $\kappa_{t}$ can be taken to be deterministic.
		
		%\item
		%Like in Remark~2.3 of \citet{carassusrasonyi2012},
		We notice that the above `quantitative' characterisation of \eqref{eq:NA} holds true only for $\mathscr{F}_{t-1}$-measurable, $\mathbb{R}^{d}$-valued functions $\xi$ which belong to $D_{t}$ a.s. This will motivate the use of orthogonal projections later on (cf.\ Section~\ref{sec:OneStep}).
	%\end{enumerate}
\end{remark}

%=========================================================================================================================================
%
% The investor
%
%=========================================================================================================================================

\subsection{The investor}

%As stated in the Introduction, %
Investors' risk preferences are described by a (possibly non-concave and non-differentiable random) utility function.%, which must satisfy certain properties with well-established economic motivations.
\begin{definition}[\textbf{Non-concave random utility}]
	A \emph{random utility} (on the non-negative half-line) is any function $u:~\left(0,+\infty\right)\times\Omega\rightarrow\mathbb{R}$ verifying the following two properties,
	\begin{enumerate}[label=\emph{(\roman*)}]
		\item
		for every $x\in\left(0,+\infty\right)$, the function $u\!\left(x,\cdot\right):~\Omega\rightarrow\mathbb{R}$ is $\mathscr{F}$-measurable,
		
		\item\label{item:DefRanUtII}
		for %$\mathbb{P}$-a.e.\ %
		a.e. $\omega\in\Omega$, the function $u\!\left(\cdot,\omega\right):~\left(0,+\infty\right)\rightarrow\mathbb{R}$ is non-decreasing and continuous.

%, and moreover its effective domain, given by
		%\begin{equation}
		%	\dom u\!\left(\cdot,\omega\right)\triangleq\left\{x\in\left[\left.0,+\infty\right)\right.\text{: }u\!\left(x,\omega\right)>-\infty\right\},
		%\end{equation}
		%is a non-empty subset of $\left[\left.0,+\infty\right)\right.$.
	\end{enumerate}
\end{definition}
For each $\omega\in\Omega$ for which $\ref{item:DefRanUtII}$ holds, we set
$u\!\left(0,\omega\right)\triangleq\lim_{x\downarrow 0}u\!\left(x,\omega\right)$, and we define
$u(0,\omega)\triangleq0$ otherwise. Note that $u\!\left(0,\omega\right)$ may take the value $-\infty$.

\begin{remark}
%	\begin{enumerate}[label=\emph{(\roman*)}]
%		\item
		As in this paper we restrict wealth to be non-negative, we consider utilities which are defined
		only over the non-negative real line. Continuity and monotonicity are standard assumptions.
		Also, as $u$ will be used to assess the future wealth of the investor, it may well depend on economic variables and hence it can be random,
		see Example \ref{ex:refpoint} below. %
%		
		%\item
		%The effective domain of $u$, $\dom u\triangleq\left\{x\in\left[\left.0,+\infty\right)\right.\text{: }u\!\left(x\right)>-\infty\right\}$, contains the entire semi-infinite interval $\left(0,+\infty\right)$.
%		
		%\item
		%We do not make any assumption concerning the differentiability of $u$. More importantly, we do not require that a utility function should be concave.
		Lastly, unlike most studies, we do not assume concavity or smoothness of $u$. A possible extension to $u$ which is only upper semicontinuous will be subject of future research.
%	\end{enumerate}
\end{remark}

%We end this subsection %
We proceed %
by noticing that, since $u\!\left(\cdot,\omega\right)$ is a monotone function for a.e.\ $\omega\in\Omega$, the limit %
%\begin{equation*}
	$u\!\left(+\infty,\cdot\right)\triangleq \lim_{x \rightarrow +\infty} u\!\left(x,\cdot\right)$ %
%\end{equation*}
exists a.s.\ (though it may not be finite), and we define $u(+\infty,\omega)\triangleq+\infty$ otherwise.
We shall require the following.
\begin{assumption}\label{as:u0}
	The negative part of $u$ at $0$ has finite expectation, that is,\footnote{Here $x^{+}\triangleq\max\left\{x,0\right\}$ and $x^{-}\triangleq-\min\left\{x,0\right\}$ for every $x\in\mathbb{R}$. Furthermore, in order to make the notation less heavy, given any function $f:~X\rightarrow\mathbb{R}$, we shall write henceforth $f^{\pm}\!\left(x\right)\triangleq \left[f\!\left(x\right)\right]^{\pm}$ for all $x\in X$.%
	}
	\begin{equation}
		\mathbb{E}_{\mathbb{P}}\!\left[u^{-}\!\left(0,\cdot\right)\right]<+\infty.
	\end{equation}
\end{assumption}
%In the sequel we shall often omit the dependence of $u$ on $\omega$ in the notation.

\begin{remark}
	%Note that, under the assumption above, we have by path monotonicity that $\dom u\!\left(\cdot,\omega\right)=\left[\left.0,+\infty\right)\right.$ for $\mathbb{P}$-a.e.\ $\omega\in\Omega$. We also recall that this assumption is not required in the treatment of the concave case given by \citet{rsch06}. Even though we are aware that this is quite restrictive a condition (for example, it excludes the widely employed utility $u\!\left(x,\omega\right)\triangleq \log\!\left(x\right)$, for all $x\geq 0$ and $\omega\in\Omega$), we do not see how to avoid it in the present setting.
	If $u$ is deterministic, then the above assumption is equivalent
	to $u\!\left(0\right)>-\infty$. This is admittedly restrictive, as it excludes
	that $u\!\left(x\right)$ behaves like $\ln\!\left(x\right)$ or $-x^{\alpha}$ (with $\alpha<0$) in the vicinity of $0$. It still allows, however, a very large class of utilities.
For the moment we cannot dispense with this hypothesis as it is crucial in proving that dynamic
programming preserves the growth condition \eqref{eq:lemgrowthu} below (see part \ref{item:3.3.v'} in the proof of
Theorem \ref{th:optstrategy} in Section \ref{sec:Dynamic}). Note also that this assumption is the pendant of (10) in Assumption 2.9 in \citet{carassusrasonyi2012} when the domain of the utility function is equal to the whole real line.

%We also draw attention to the fact that, in the concave case (see \citet{rsch06}), such an assumption was
	%not needed. Nonetheless, we do not see how to avoid it in the present setting. See examples.
	%{\todo{HEREEE. What examples?}}
\end{remark}

We continue this subsection with an important example of a random utility function, in the spirit
of cumulative prospect theory.%(CPT, see \citet{kahneman79,tversky92}).
\begin{example}[Reference point]\label{ex:refpoint}
	Within the CPT framework, every investor is assumed to have a \emph{reference point} in wealth
	(also referred to  as \emph{benchmark} or \emph{status quo} in the literature, see e.g.\ \citet{bernard2010}, \citet{he2011}, \citet{carassusrasonyi2011}), with respect to which payoffs at the terminal time $T$ are evaluated. Therefore, investors' decisions are not based on the terminal level of wealth (as it is assumed in the Expected Utility Theory of \citet{neumann53}), but rather on the deviation of that wealth level from the reference point. Note that, unlike in CPT, our setting does not include probability distortions
	(weight functions).
	
	Mathematically, a reference point is any fixed scalar-valued and
	$\mathscr{F}$-measurable random variable $B\geq 0$ a.s. %\ (we recall that we do not allow wealth to become strictly negative). %
	%with $\esssup B<+\infty$.
	Thus, given a payoff $X$ at the terminal time $T$ and a scenario $\omega\in\Omega$, the investor is said to make a \emph{gain} (respectively, a \emph{loss}) if the deviation from the reference level is strictly positive (respectively, strictly negative), that is, $X\!\left(\omega\right)>B\!\left(\omega\right)$ (respectively, $X\!\left(\omega\right)<B\!\left(\omega\right)$).
	
	Note that $B$ may be taken to be, for example, a non-negative constant %$x_{0}$, that is, the terminal wealth of the portfolio consisting of investing all of the initial wealth $x_{0}$ in the riskless asset %
	(this is the case in \citet{bernard2010,berkelaar04,carassus-pham}). The reference point can also be stochastic (for instance, to reflect the fact that the investors compare their performance to that of another investor acting in a perhaps different market).
	
	In this setting, the investor has a random utility defined as
	\begin{equation}
		u\!\left(x,\omega\right)\triangleq \widetilde{u}\!\left(x-B\!\left(\omega\right)\right),\quad x>0,\ \omega\in\Omega,
	\end{equation}
	with $\widetilde{u}:~\left(-\esssup B,+\infty\right)\rightarrow\mathbb{R}$ a (deterministic) non-decreasing and continuous function satisfying
	$\widetilde{u}\!\left(-\esssup B\right)>-\infty$ (where we set $\widetilde{u}\!\left(-\esssup B\right)\triangleq \lim_{x\downarrow-\esssup B}\widetilde{u}\!\left(x\right)$, as before). Obviously, %
	%for $\mathbb{P}$-a.e. $\omega\in\Omega$, we have $u\!\left(0,\omega\right)=\lim_{x\downarrow0}\widetilde{u}\!\left(x-B\!\left(\omega\right)\right)\geq\lim_{x\downarrow0}\widetilde{u}\!\left(x-\esssup B\right)=\widetilde{u}\!\left(-\esssup B\right)$, which implies that %
	$\mathbb{E}_{\mathbb{P}}\!\left[u^{-}\!\left(0,\cdot\right)\right]<+\infty$, so Assumption~\ref{as:u0} is true for $u$.
\end{example}

We shall make the following assumption on the growth of the function $u$.
\begin{assumption}\label{as:growthu}
	There exist constants $\overline{\gamma}>0$ and $\overline{x}\geq 0$,
	as well as a random variable $c\geq 0$ a.s.
with $\mathbb{E}_{\mathbb{P}}\!\left[c\right]<+\infty$, %
	such that for a.e. $\omega\in\Omega$,
	\begin{equation}\label{eq:asgrowthu}
		%u\!\left(\lambda x\right)\leq \lambda^{\overline{\gamma}}u\!\left(x\right)+\lambda^{\overline{\gamma}}c,
		u\!\left(\lambda x,\omega\right)\leq \lambda^{\overline{\gamma}}u\!\left(x,\omega\right)+\lambda^{\overline{\gamma}}c\!\left(\omega\right)
	\end{equation}
holds simultaneously for all $\lambda\geq 1$ and for all $x\geq \overline{x}$. Furthermore,
$\mathbb{E}_{\mathbb{P}}\left[u^+(\overline{x},\cdot)\right]<+\infty$.
\end{assumption}

\begin{remark}\label{obs:growthu}
	For $u$ deterministic, strictly concave and continuously differentiable, we recall that
	\begin{equation*}
		AE_{+}\!\left(u\right)\triangleq \limsup_{x\rightarrow+\infty}\frac{x\,u'\!\left(x\right)}{u\!\left(x\right)}
	\end{equation*}
	denotes the \emph{asymptotic elasticity} of $u$ at $+\infty$ (see \citet[p.~943]{kramkov99}), and
	we always have $AE_{+}\!\left(u\right)\leq 1$ (the reader is referred to \citet[Lemma~6.1]{kramkov99}).
	
%	If, in addition, $u\!\left(+\infty\right)>0$,
We know by Lemma~6.3 in \citet{kramkov99} that %$AE_{+}\!\left(u\right)\leq\gamma$ (with $\gamma>0$) is equivalent to (actually, it is necessary, but not sufficient; $AE_{+}\!\left(u\right)<\gamma$ is sufficient) the	 existence of some $\overline{x}>0$ such that, for all $\lambda\geq 1$ and for all $x\geq \overline{x}$,
	$AE_{+}\!\left(u\right)$ equals the infimum of all real numbers $\gamma>0$ for which there exists some $\overline{x}>0$ such that, for all $\lambda\geq 1$ and all $x\geq \overline{x}$,
	\begin{equation}\label{planetarium}
		u(\lambda x)\leq \lambda^{\gamma}u\!\left(x\right)
	\end{equation}%
holds.\footnote{To be precise, in the cited lemma there is strict inequality in \eqref{planetarium} and it
is required to hold for $\lambda>1$ only. As easily seen, it works also for our version.}
%under $u\!\left(+\infty\right)>0$ this is equivalent to our version of \eqref{planetarium} here.}
From the proof of Lemma~6.3 of \citet{kramkov99}, it is clear that this characterisation of $AE_+(u)$ also holds true if $u$ is not concave (but continuously differentiable). As this latter formulation \eqref{planetarium} makes sense for possibly non-differentiable and non-concave $u$ and arbitrary $\gamma>0$ as well, we follow \citet{carassusrasonyi2012} and define the asymptotic elasticity at $+\infty$ of $u$ as
	\begin{equation*}
		AE_{+}\!\left(u\right)\triangleq \inf\!\left\{\gamma>0\text{ : }\exists\,\overline{x}\geq0\text{ s.t. for a.e. }\omega,
		%\mathbb{E}_{\mathbb{P}}\left[u^+(\overline{x},\cdot)\right]<\infty  \text{ and }
		u\!\left(\lambda x,\omega\right)\leq\lambda^{\gamma}u\!\left(x,\omega\right),\ \forall\,\lambda\geq1,\ \forall\,x\geq\overline{x}\right\},
	\end{equation*}
	with the usual convention that the infimum of the empty set is $+\infty$.
	
	Hence, using this generalized notion of asymptotic elasticity, we see that condition~\eqref{eq:asgrowthu} holds if either $AE_{+}\!\left(u\right)<+\infty$, or $u$ is  bounded above by some integrable random constant $C\geq0$ a.s.\ and Assumption \ref{as:u0} holds. %
	 %after some constant $\overline{x} >0$  such that
	%$\mathbb{E}_{\mathbb{P}}\left[u^+(\overline{x},\cdot)\right]<\infty$.
	Indeed, in the first case this is trivial since \eqref{eq:asgrowthu} is implied by $AE_{+}\!\left(u\right)<\overline{\gamma}$. In the second case, taking $\overline{x}\triangleq 0$, we get that for every $x \geq \overline{x}$ and for every $\lambda \geq 1$,
\begin{align*}
	u\!\left(\lambda x,\omega\right)\leq C\!\left(\omega\right) \leq \lambda^{\overline{\gamma}}C\!\left(\omega\right)&=
\lambda^{\overline{\gamma}}u\!\left(x,\omega\right)+\lambda^{\overline{\gamma}}
\left[C(\omega)-u\!\left(x,\omega\right)\right]\\
	&\leq\lambda^{\overline{\gamma}}u\!\left(x,\omega\right)+\lambda^{\overline{\gamma}}
\left[C(\omega)-u\!\left(0,\omega\right)\right],
\end{align*}
which permits us to define $c\!\left(\omega\right)\triangleq\left[C\!\left(\omega\right)-u\!\left(0,\omega\right)\right]^{+}$.
As $c(\omega) \leq C\!\left(\omega\right)+u^-\!\left(0,\omega\right)$, Assumption \ref{as:u0} shows that
$\mathbb{E}_{\mathbb{P}}\left[c\right] <+\infty$.
Note that in this case $\mathbb{E}_{\mathbb{P}}\left[u^+(\overline{x},\cdot)\right]=\mathbb{E}_{\mathbb{P}}\left[u^+(0,\cdot)\right] \leq
\mathbb{E}_{\mathbb{P}}[C]<+\infty$

%\lambda^{\overline{\gamma}}u\!\left(x,\omega\right)
%+ \lambda^{\overline{\gamma}}\left(2 \times C(\omega)\right).$$
	
	We note that, if $u$ is deterministic, concave and bounded above, then $AE_{+}\!\left(u\right)\leq 0$ (again by \citet[Lemma~6.1]{kramkov99}), but this fails in the non-concave case. As we will see, in Example~\ref{ex:boundedAEinf} below, Assumption~\ref{as:growthu} holds true, but the asymptotic elasticity is equal to $+\infty$. This shows that having finite asymptotic elasticity, despite being sufficient, is not a necessary condition for a function to verify Assumption~\ref{as:growthu}.

	We immediately get that $AE_{+}\!\left(u\right)<+\infty$ (and hence Assumption~\ref{as:growthu} holds) provided that
	$u$ is deterministic, continuously differentiable and there exists some $p>0$ such that
	\begin{equation*}
		0<\liminf_{x\rightarrow+\infty} \frac{u'\!\left(x\right)}{x^{p}}\leq\limsup_{x\rightarrow+\infty} \frac{u'\!\left(x\right)}{x^{p}}<+\infty.
	\end{equation*}
	%(namely, if $u'\!\left(x\right)$ is asymptotically equivalent to $x^{p}$ as $x\rightarrow+\infty$). %
	Indeed, if the above condition is true for $u$, then on the one hand it is possible to find
	$m>0$ for which there exists some $\underline{x}>0$ such that
	$u'\!\left(x\right)>m\,x^{p}$ for all $x\geq\underline{x}$. But this implies that,
	for all $x\geq\underline{x}$,
	\begin{equation*}
		u\!\left(x\right)-u\!\left(\underline{x}\right)= \int_{\underline{x}}^{x}u'\!\left(y\right)\,dy\geq m\,\frac{x^{p+1}-\underline{x}^{p+1}}{p+1}.
	\end{equation*}
	On the other hand, we can find $M>0$ for which there is $x_{M}>0$ such that $u'\!\left(x\right)<M\,x^{p}$ for all $x\geq x_{M}$. Defining $\check{x}\triangleq\max\left\{\underline{x},x_{M}\right\}>0$, noticing that we may assume that $u\!\left(\underline{x}\right)>0$ without loss of generality, and combining the preceding inequalities finally gives
	\begin{equation*}
		\frac{x\,u'\!\left(x\right)}{u\!\left(x\right)}\leq \frac{Mx^{p+1}}{\frac{m}{p+1}\,\left(x^{p+1}-\underline{x}^{p+1}\right) + u\!\left(\underline{x}\right)}
	\end{equation*}
%\begin{equation*}
%		\frac{x\,u'\!\left(x\right)}{u\!\left(x\right)}\leq \left(p+1\right)\frac{M}{m}\,\frac{x^{p+1}}{x^{p+1}-\underline{x}^{p+1}}
%	\end{equation*}
	for all $x\geq \check{x}$% large enough (such that the denominator is non-zero)%
	, therefore %$\limsup_{x\rightarrow+\infty}x\,u'\!\left(x\right)/u\!\left(x\right)<+\infty$.
	\begin{equation*}
		\limsup_{x\rightarrow+\infty}\frac{x\,u'\!\left(x\right)}{u\!\left(x\right)}<+\infty.
	\end{equation*}
	In particular, %
	%if $u(x)$ equals %
	if $u'\!\left(x\right)$ is asymptotically equivalent to %
	a power function (that is, $u'(x)/x^p\to 1$, as $x\to+\infty$) %
	then Assumption~\ref{as:growthu} holds. A multitude of piecewise concave or ``\emph{S}-shaped'' functions (not only
	piecewise power functions) can be accomodated in this way,
	such as the ones considered in \citet{berkelaar04,jin2008}.
	
	At last, suppose that $u$ is the utility of Example~\ref{ex:refpoint}. If the conditions below are satisfied:
	\begin{enumerate}[label=\emph{(\roman*)}]
		\item
		$\esssup B<+\infty$;
		
		\item
		there exist real numbers $\overline{\gamma}>0$, $\widetilde{x}>0$ and $C\geq 0$ such that, for all $\lambda\geq 1$ and all $x\geq\widetilde{x}$,
		\begin{equation*}
			\widetilde{u}\!\left(\lambda x\right)\leq \lambda^{\overline{\gamma}}\widetilde{u}\!\left(x\right)+\lambda^{\overline{\gamma}}C;
		\end{equation*}
		
		\item
		the function $\widetilde{u}$ is continuously differentiable on its domain, and there are real numbers $K>0$ and $\widehat{x}>0$ such that, for all $x\geq \widehat{x}$,
		\begin{equation*}
			\widetilde{u}'\!\left(x\right)\leq K;
		\end{equation*}
	\end{enumerate}
then $u$ fullfills Assumption~\ref{as:growthu}. Indeed,
	setting $\overline{x}\triangleq \max\!\left\{\widetilde{x},\widehat{x}\right\}+\esssup B>0$ yields
	\begin{align*}
		u\!\left(\lambda x,\omega\right)=\widetilde{u}\!\left(\lambda\left[x-\frac{B\!\left(\omega\right)}{\lambda}\right]\right)&\leq \lambda^{\overline{\gamma}}\widetilde{u}\!\left(x-\frac{B\!\left(\omega\right)}{\lambda}\right)+\lambda^{\overline{\gamma}}C\\
		&\leq \lambda^{\overline{\gamma}}\widetilde{u}\!\left(x-B\!\left(\omega\right)\right)+\lambda^{\overline{\gamma}}K\,B\!\left(\omega\right)\left(1-\frac{1}{\lambda}\right)+\lambda^{\overline{\gamma}}C
	\end{align*}
	for a.e.\ $\omega$, simultaneously for all $\lambda\geq 1$ and $x\geq \overline{x}$.
	Note that $u^+(\overline{x},\cdot)\leq
	\tilde{u}^+(\overline{x})$  and the latter is deterministic.
	Hence, choosing $c\triangleq K\,\esssup B+C$
	(which is constant, thus trivially integrable) gives the claimed result.
	We conclude by pointing out that any funcion $\widetilde{u}$ which is concave for sufficiently large $x$
	satisfies the conditions \emph{(ii)}, \emph{(iii)} above.
\end{remark}

We may now deduce the following auxiliary result, which provides an estimate for all $x\geq 0$, and not only for $x\geq\overline{x}$.
\begin{lemma}\label{lem:growthu}
	Under Assumption~\ref{as:growthu} there is a random variable $C\geq 0$ a.s.\ such that $\mathbb{E}_{\mathbb{P}}\!\left[C\right]<+\infty$ and, for a.e.\ $\omega$,
	\begin{equation}\label{eq:lemgrowthu}
		u^{+}\!\left(\lambda x,\omega\right)\leq \lambda^{\overline{\gamma}} u^{+}\!\left(x,\omega\right)+\lambda^{\overline{\gamma}}C\!\left(\omega\right)
	\end{equation}
	simultaneously for all $\lambda\geq 1$ and for all $x\geq0$.
	\end{lemma}

\begin{proof}
	See Appendix~\ref{sec:Aux}.%, \cpageref{proof:lem:growthu}.
\end{proof}

%=========================================================================================================================================
%
% The optimal portfolio problem
%
%=========================================================================================================================================

\section{Main results}\label{ExistTheo}
%\subsection{Existence theorem}\label[subsec]{ExistTheo}

%As already described in the Introduction, %
The optimal portfolio problem consists in choosing the ``best'' investment in the given assets: the one
which maximises the expected utility from terminal wealth. %In other words, since we are assuming that there is no intermediate consumption, the investors with a given non-negative initial capital $x_{0}$ wish to select, from all allowable portfolios, the investment strategies whose terminal wealths give them the highest expected utility.

%This can be mathematically formalised as follows.%NOTE: we need the following definition BEFORE the assumption that u(1)=0
\begin{definition}%[\textbf{Non-concave portfolio choice problem}]
	Let Assumption~\ref{as:u0} be in force. Given any $x_{0}\geq 0$, the \emph{non-concave portfolio problem} with initial wealth $x_{0}$ on a finite horizon $T$
	is to find $\phi^*\in\Psi(x_0)$ such that
	\begin{equation}%\tag{NCPP}
		v^{*}\!\left(x_{0}\right)\triangleq\sup\!\left\{\mathbb{E}_{\mathbb{P}}\left[u\!\left(\Pi^{\phi}_{T}(\cdot),\cdot\right)\right]\text{: }\phi\in\Psi\!\left(x_{0}\right)\right\}
		=\mathbb{E}_{\mathbb{P}}\left[u\!\left(\Pi^{\phi^{*}}_{T}(\cdot),\cdot\right)\right].
		\label{eq:OP}
	\end{equation}
	%\begin{equation*}
	%	\sup\!\left\{\mathbb{E}_{\mathbb{P}}\left[u\!\left(\Pi^{\phi}_{T}\!\left(\cdot\right),\cdot\right)\right]\text{: }\phi\in\mathscr{A}\!\left(x_{0}\right)\right\},
	%\end{equation*}
	%where the \emph{feasible set} is given by
	%\begin{equation}
	%	\mathscr{A}\!\left(x_{0}\right)\triangleq \left\{\phi\in\Psi\!\left(x_{0}\right)\text{: }\mathbb{E}_{\mathbb{P}}\left[u^{+}\!\left(\Pi^{\phi}_{T}\!\left(\cdot\right),\cdot\right)\right]<+\infty\text{ or }\mathbb{E}_{\mathbb{P}}\left[u^{-}\!\left(\Pi^{\phi}_{T}\!\left(\cdot\right),\cdot\right)\right]<+\infty\right\}.
	%\end{equation}
	%Setting $ \sup\!\left\{\mathbb{E}_{\mathbb{P}}\left[u\!\left(\Pi^{\phi}_{T}\right)\right]\text{: }\phi\in\Psi\!\left(x_{0}\right)\right\}$, we say that $\phi^{*}\in\Psi\!\left(x_{0}\right)$ is an \emph{optimal strategy} if
	%\begin{equation}
	%	v^{*}\!\left(x_{0}\right)=.
	%\end{equation}
	%Setting $v^{*}\!\left(x_{0}\right)\triangleq \sup\!\left\{\mathbb{E}_{\mathbb{P}}\left[u\!\left(\Pi^{\phi}_{T}\!\left(\cdot\right),\cdot\right)\right]\text{: }\phi\in\mathscr{A}\!\left(x_{0}\right)\right\}$, we say that $\phi^{*}\in\mathscr{A}\!\left(x_{0}\right)$ is an \emph{optimal strategy} if
	%\begin{equation}
	%	v^{*}\!\left(x_{0}\right)=\mathbb{E}_{\mathbb{P}}\left[u\!\left(\Pi^{\phi^{*}}_{T}\!\left(\cdot\right),\cdot\right)\right].
	%\end{equation}
	We call $\phi^*$ an \emph{optimal strategy}.
	\end{definition}
%Note that, due to Assumption~\ref{as:u0}, the expectations in \eqref{eq:OP} above exist, though they may be infinite.

\begin{remark}\label{obs:significance}
	\begin{enumerate}[label=\emph{(\roman*)}]
	
	\item\label{obs:vstargequx0}
	Note that, due to Assumption~\ref{as:u0}, %
	\begin{equation*}
		 \mathbb{E}_{\mathbb{P}}\!\left[u^{-}\!\left(\Pi^{\phi}_{T}(\cdot),\cdot\right)\right]\leq\mathbb{E}_{\mathbb{P}}\!\left[u^{-}\!\left(0,\cdot\right)\right]<+\infty,
	\end{equation*}
	and the expectations in \eqref{eq:OP} above exist (though they may be infinite). It is also 	immediate to check that, %the feasible set is non-empty. Indeed, for every $x_{0}\in\left[\left.0,+\infty\right)\right.$, the trivial portfolio $\overline{\varphi}_{x_{0}}$ given by
	%\begin{equation*}\label{eq:discretetrivialportfolio}
	%	\left(\varphi_{x_{0}}\right)^{0}_{t}\!\left(\omega\right)\triangleq x_{0} \quad \text{and}\quad \left(\varphi_{x_{0}}\right)^{i}_{t}\!\left(\omega\right)\triangleq 0,\quad \forall\,\omega\in\Omega,\ \forall\,t\in\left\{1,\ldots,T\right\},\ \forall\,i\in\left\{1,\ldots,d\right\},
	%\end{equation*}
	%that is, consisting in investing all of the wealth on the bond and none on the risky assets, belongs to $\mathscr{A}\!\left(x_{0}\right)$. %
	the strategy $\phi\equiv 0$ is in $\Psi\!\left(x_{0}\right)$ for all $x_{0}\geq 0$, so the supremum is taken over a non-empty set. %
	%This implies %
	In particular, %that %
	$v^{*}\!\left(x_{0}\right)\geq \mathbb{E}_{\mathbb{P}}\!\left[u\!\left(x_{0},\cdot\right)\right]>-\infty$, under Assumption~\ref{as:u0}.

	\item
	One may inquire why the existence of an optimal $\phi^{*}$ is important
	when the existence of $\varepsilon$-optimal strategies $\phi^{\varepsilon}$ (i.e., ones that are $\varepsilon$-close to the supremum over all strategies) is automatic, for all $\varepsilon>0$.
	
	Firstly, non-existence of an optimal strategy $\phi^{*}$ usually means
	that an optimiser sequence $\left\{\phi^{1/n}\text{; }n\in\mathbb{N}\right\}$ shows
	a behaviour which is practically infeasible and counter-intuitive (see Example~7.3 of \citet{rs05}).
	
	Secondly, existence of $\phi^{*}$ normally goes together with some compactness property
	which would be needed for the construction of eventual numerical schemes to find the optimiser.
	%Such a property seems necessary for the convergence of any potential numerical
	%procedure to find an optimal (or at least an $\varepsilon$-optimal) strategy.
	\end{enumerate}
\end{remark}

%We conclude this %section %
%subsection %
%with the following result, which says that, if we wish to have any hope of finding an optimal portfolio, then \eqref{eq:NA} cannot be dropped when the utility $u$ is assumed to be strictly increasing.
%\begin{proposition}\label[prop]{prop:NAnec}
%	Let $x_{0}\in\left[\left.0,+\infty\right)\right.$	be arbitrary. Suppose that $u$ is strictly increasing, and that $v^{*}\!\left(x_{0}\right)<+\infty$. Then, there exists an optimal portfolio $\phi^{*}\in\mathscr{A}\!\left(x_{0}\right)$ for \eqref{eq:OP} only if \eqref{eq:NA} holds true.
%\end{proposition}

%\begin{proof}
%	This is a restatement of Proposition~3.1 in \citet{rs05}.
%\end{proof}

%%%%%%%%%%%%%%%%%%%%%%%%%%%%%%%%%%%%%%%%%%%%%%%%%%%%%%%%%%%%%%%%%%%%%%%%%%%%%%%%%%%%%%%%%%%%%%%%%%%%%%%%%%%%%%%%%%%%%%%%%%%%%%%%%%%%%%%%%%
%%                                                                                                                                      %%
%% Existence theorem                                                                                                                    %%
%%                                                                                                                                      %%
%%%%%%%%%%%%%%%%%%%%%%%%%%%%%%%%%%%%%%%%%%%%%%%%%%%%%%%%%%%%%%%%%%%%%%%%%%%%%%%%%%%%%%%%%%%%%%%%%%%%%%%%%%%%%%%%%%%%%%%%%%%%%%%%%%%%%%%%%%

Here comes the main result of the present paper. It says that the optimisation problem \eqref{eq:OP}
admits a solution.
\begin{theorem}\label{th:optstrategy}
	Let Assumptions~\ref{as:NA}, \ref{as:u0} and \ref{as:growthu} hold true. Assume further that, for every $x_{0}\in\left[\left.0,+\infty\right)\right.$,
	\begin{equation}%\label{eq:supEfinite}
		%\sup_{\phi\in\Psi\!\left(x_{0}\right)}\mathbb{E}_{\mathbb{P}}\!\left[u\!\left(\Pi^{\phi}_{T}\right)\right]<+\infty.
		v^{*}\!\left(x_{0}\right)<+\infty.
		\label{eq:supEfinite}
	\end{equation}
	Then, for each $x_{0}\in\left[\left.0,+\infty\right)\right.$, there exists a strategy $\phi^{*}\in\Psi\!\left(x_{0}\right)$ satisfying
	\begin{equation}
		\mathbb{E}_{\mathbb{P}}\!\left[u\!\left(\Pi^{\phi^{*}}_{T}(\cdot),\cdot\right)\right]=v^{*}\!\left(x_{0}\right).
	\end{equation}
\end{theorem}

\begin{proof}
The proof will be given in Section \ref{sec:Dynamic}, after appropriate preparations. We give here a brief description. A dynamic programming technique will be applied. This will allow us to split the original problem into several sub-problems involving
a random utility function $U_{t}$ at time $t$. At each time step, we will find a one-step optimal solution based on the natural compactness provided by Lemma \ref{lem:rvKx} below and on the random subsequence technique of \citet{kabanov01} (see Sections~\ref{sec:OneStep} and \ref{sec:Aux}). Furthermore, we will prove that certain crucial properties of $U_{t}$,
such as continuity and the growth condition (\ref{eq:lemgrowthu}), are preserved for the next iteration (i.e. for $U_{t-1}$). These are the most involved arguments of the present paper. Finally, we shall paste together the one-step maximisers in a natural way to get a maximiser in \eqref{eq:supEfinite}.
\end{proof}

\begin{remark}
	We would like to stress that, since Assumption~\ref{as:u0} is in force, the well-posedness condition \eqref{eq:supEfinite} is actually equivalent to the apparently stronger one

\begin{equation}\label{lepra}
		\sup_{\phi\in\Psi\!\left(x_{0}\right)}\mathbb{E}_{\mathbb{P}}\!\left[u^{+}\!\left(\Pi^{\phi}_{T}(\cdot),\cdot\right)\right]<+\infty.
\end{equation}
	To see this, we recall that $\Pi^{\phi}_{T}\geq 0$ a.s.\ for every $\phi\in\Psi\!\left(x_{0}\right)$,
	hence
	\begin{equation*}
		\sup_{\phi\in\Psi\!\left(x_{0}\right)}\mathbb{E}_{\mathbb{P}}\!\left[u^{+}\!\left(\Pi^{\phi}_{T}(\cdot),\cdot\right)\right]\leq v^{*}\!\left(x_{0}\right)+\sup_{\phi\in\Psi\!\left(x_{0}\right)}\mathbb{E}_{\mathbb{P}}\!\left[u^{-}\!\left(\Pi^{\phi}_{T}(\cdot),\cdot\right)\right]\leq v^{*}\!\left(x_{0}\right)+\mathbb{E}_{\mathbb{P}}\!\left[u^{-}\!\left(0,\cdot\right)\right]<+\infty.
	\end{equation*}	

\end{remark}

As a very simple, yet important example to which the preceding theorem clearly applies, we mention the case
where $S$ satisfies
Assumption~\ref{as:NA} and
$u(x,\omega)$  is bounded above by some integrable random constant $C(\omega)$, for all $x$, and satisfies Assumption \ref{as:u0}. Another relevant example is given
by the following theorem. First, define
	\begin{equation}\label{eq:defWfamily}
		\mathscr{W}\triangleq \left\{Y\in\Xi^{1}_{T}\text{: }\mathbb{E}_{\mathbb{P}}\!\left[\left|Y\right|^{p}\right]<+\infty \text{ for all }p>0\right\}.
	\end{equation}%

\begin{theorem}\label{th:StinW}
	Let Assumptions~\ref{as:NA}, \ref{as:u0} and \ref{as:growthu} hold true with $c,u^+(\overline{x},\cdot)\in\mathscr{W}$.
	Assume further that $\left\Vert\Delta S_{t}\right\Vert,1/\beta_{t}\in\mathscr{W}$ for every
	$t\in\left\{1,\ldots,T\right\}$, where the $\beta_{t}$ are the random variables figuring in
	Proposition~\ref{prop:quantNA}.
	Then, for every $x_{0}\in\left[\left.0,+\infty\right)\right.$, %
	condition~\eqref{eq:supEfinite} is satisfied and %
	there exists an optimal strategy $\phi^{*}\in\Psi\!\left(x_{0}\right)$.
\end{theorem}

\begin{proof} See Section~\ref{sec:Dynamic}.
%	Proposition~5.1 in \citet{rsch06} shows that there exists some $X\in\mathscr{W}$ such that, for all $\phi\in\Psi\!\left(x_{0}\right)$, the inequality $\Pi^{\phi}_{T}\leq X$ holds a.s. Then Assumption~\ref{as:growthu} shows that \ldots,	hence $\mathbb{E}_{\mathbb{P}}\!\left[u\!\left(\Pi^{\phi}_{T}\right)\right]\leq \mathbb{E}_{\mathbb{P}}\!\left[X^q\right]$ for $q=\ldots$, and this is clearly finite.{\todo{HEREEE. Write.}}
\end{proof}

%%%%%%%%%%%%%%%%%%%%%%%%%%%%%%%%%%%%%%%%%%%%%%%%%%%%%%%%%%%%%%%%%%%%%%%%%%%%%%%%%%%%%%%%%%%%%%%%%%%%%%%%%%%%%%%%%%%%%%%%%%%%%%%%%%%%%%%%%%
%%                                                                                                                                      %%
%% The one-step case                                                                                                                    %%
%%                                                                                                                                      %%
%%%%%%%%%%%%%%%%%%%%%%%%%%%%%%%%%%%%%%%%%%%%%%%%%%%%%%%%%%%%%%%%%%%%%%%%%%%%%%%%%%%%%%%%%%%%%%%%%%%%%%%%%%%%%%%%%%%%%%%%%%%%%%%%%%%%%%%%%%

\section{The one-step case}\label{sec:OneStep}

In this section, we consider an $\mathscr{F}$-measurable function $Y:~\Omega\rightarrow \mathbb{R}^{d}$, and a $\sigma$-algebra $\mathscr{G}\subseteq\mathscr{F}$ containing all $\mathbb{P}$-null sets of $\mathscr{F}$. This setting will be applied in the multi-step case (see the subsequent section) with $\mathscr{G}=\mathscr{F}_{t-1}$ and $Y=\Delta S_{t}$, for every fixed $t\in\left\{1,\ldots,T\right\}$.

Keeping in line with the notation of the previous section, we denote by $\Xi^{d}$ the family of all $\mathscr{G}$-measurable functions $\xi:~\Omega\rightarrow\mathbb{R}^{d}$.

Moreover, let $\rcd{Y}{G}:~\mathscr{B}\!\left(\mathbb{R}^{d}\right)\times\Omega\rightarrow\left[0,1\right]$ be the
unique (up to a set of measure zero) regular conditional distribution for $Y$ given $\mathscr{G}$.
By modifying it on a $\mathbb{P}$-null set, we may and will assume that $\rcd{Y}{G}(\cdot,\omega)$
is a probability for \emph{each} $\omega$.
Now, for each $\omega\in\Omega$,
let $\supp\!\left(\rcd{Y}{G}\!\left(\cdot,\omega\right)\right)$ represent the support of $\rcd{Y}{G}\!\left(\cdot,\omega\right)$ (which exists and is non-empty), %, see \Cref{prop:existsupp}), %
and let $D\!\left(\omega\right)$ denote the affine hull of $\supp\!\left(\rcd{Y}{G}\!\left(\cdot,\omega\right)\right)$, that is, %
%\begin{equation*}
$	D\!\left(\omega\right)\triangleq \aff\!\left(\supp\!\left(\rcd{Y}{G}\!\left(\cdot,\omega\right)\right)\right)$. %
%\end{equation*}

We shall also assume the following.
\begin{assumption}\label{as:Dlinsub}
 For a.e. $\omega\in{\Omega}$, $D\!\left(\omega\right)$ is a linear subspace of $\mathbb{R}^{d}$.
\end{assumption}

In addition, for every $\mathscr{G}$-measurable random variable $H:~\Omega\rightarrow \mathbb{R}$ satisfying $H\geq 0$ a.s.\ (and also for any constant $x \geq 0$), define the set
\begin{equation*}
	\Xi^{d}\!\left(H\right)\triangleq \left\{\xi\in\Xi^{d}\text{: }H+\left\langle\xi,Y\right\rangle\geq 0\text{ a.s.}\right\}.
\end{equation*}
%Then in the particular case where $H=x$ a.s., for some $x\in\left[\left.0,+\infty\right)\right.$, we have
%\begin{equation*}
%	\Xi^{d}\!\left(x\right)\triangleq \left\{\xi\in\Xi^{d}\text{: }\left\langle\xi,Y\right\rangle\geq -x \text{ a.s.}\right\}.
%\end{equation*}

Finally, let $\widetilde{\Xi}^{d}$ denote the family of all functions $\xi\in\Xi^{d}$ such that %
%$\left\Vert\xi\right\Vert=1$ a.s.\ on $\left\{\omega\in\widehat{\Omega}\text{: }D\!\left(\omega\right)\neq\left\{0\right\}\right\}$ and %
$\xi\!\left(\omega\right)\in D\!\left(\omega\right)$ for a.e. $\omega$. The notation $\tilde{\Xi}^{d}\!\left(H\right)$
is  self-explanatory.

%\begin{remark}\label{obs:xi0}
%	It is trivial to see that $\widetilde{\Xi}^{d}$ is non-empty. In fact, let $\xi_{0}:\Omega\rightarrow \mathbb{R}^{d}$ be the \emph{null function}, that is, %
%	%\begin{equation*}
%	%	\xi_{0}\!\left(\omega\right)\triangleq\vect{0},\qquad \omega\in\Omega.
%	%\end{equation*}
%	$\xi_{0}\!\left(\omega\right)\triangleq\vect{0}$ for every $\omega\in\Omega$. %
%	Then we have by \Cref{as:Dlinsub} that, for every $\omega\in\widehat{\Omega}$, the affine space $D\!\left(\omega\right)$ is actually a vector space, and hence $\xi_{0}\!\left(\omega\right)=\vect{0}\in D\!\left(\omega\right)$.
%\end{remark}

We shall also impose the following condition, which can be regarded as one-step absence of arbitrage (cf.\ Proposition~\ref{prop:quantNA}).
\begin{assumption}\label{as:kappadelta}
	There exist $\mathscr{G}$-measurable random variables $\beta, \kappa>0$ a.s.\ such that
	\begin{equation}
		\mathbb{P}\!\left(\left.\left\langle \xi,Y \right\rangle\leq-\beta\left\Vert \xi \right\Vert\right|\mathscr{G}\right)\geq\kappa \text{ a.s.}
	\end{equation}
	for all $\xi\in\widetilde{\Xi}^{d}$. We may and will assume $\beta\leq 1$.
\end{assumption}

%Now let $V:~\left[\left.0,+\infty\right)\right.\times\Omega\rightarrow \mathbb{R}$ be a function verifying the following conditions.
\begin{assumption}\label{as:V}
	Let the function $V~:~\left[\left.0,+\infty\right)\right.\times\Omega\rightarrow \mathbb{R}$ satisfy
	both properties below:
	\begin{enumerate}[label=\emph{(\roman*)}]
		\item
		for any fixed $x\in\left[\left.0,+\infty\right)\right.$, the function $V\!\left(x,\cdot\right):~\Omega\rightarrow \mathbb{R}$ is measurable with respect to $\mathscr{F}$;
		\item
		for a.e. $\omega\in\Omega$, the function
		$V\!\left(\cdot,\omega\right):~\left[\left.0,+\infty\right)\right.\rightarrow \mathbb{R}$
		is continuous and non-decreasing.
	\end{enumerate}
\end{assumption}

%\begin{remark}
%	We observe that, for every $x\geq 0$ and for every $\xi\in\Xi^{d}\!\left(x\right)$, the function mapping each $\omega$ in $\Omega$ to $V\!\left(x+\left\langle \xi\!\left(\omega\right),Y\!\left(\omega\right) \right\rangle,\omega\right)$ is well-defined, except possibly on a set of $\mathbb{P}$-measure zero. From this time forth, any function on $\Omega$ which is defined for $\mathbb{P}$-a.e.\ $\omega$ is considered to be well-defined.
%\end{remark}

We shall also need the following integrability conditions.% Again we remark that the second one is rather restrictive.
\begin{assumption}\label{as:esssupfiniteas}
	For every $x\in\left[\left.0,+\infty\right)\right.$,
	\begin{equation}\label{eq:esssupfiniteas}
		\esssup_{\xi\in\Xi^{d}\!\left(x\right)} \mathbb{E}_{\mathbb{P}}\left[\left.V^{+}\!\left(x+\left\langle \xi\!\left(\cdot\right),Y\!\left(\cdot\right) \right\rangle,\cdot\right)\right|\mathscr{G}\right]<+\infty \text{ a.s.}
	\end{equation}
\end{assumption}

%\begin{remark}\label{obs:esssupfiniteas}
%	It is obvious that \Cref{as:esssupfiniteas} implies that, for all $x\geq0$,
%	\begin{equation}\label{eq:esssupV}
%		%\forall\,x\geq 0,\quad%
%		\esssup_{\xi\in\Xi^{d}\!\left(x\right)} \mathbb{E}_{\mathbb{P}}\left[\left.V\!\left(x+\left\langle \xi\!\left(\cdot\right),Y\!\left(\cdot\right) \right\rangle,\cdot\right)\right|\mathscr{G}\right]<+\infty \text{ a.s.}\qedhere
%	\end{equation}
%\end{remark}

\begin{assumption}\label{as:V-}
	The conditional expectation of $V^{-}\!\left(0,\cdot\right):~\Omega\rightarrow\left[\left.0,+\infty\right)\right.$ with respect to $\mathscr{G}$ is finite a.s., i.e.,
	\begin{equation}\label{eq:V-}
		\mathbb{E}_{\mathbb{P}}\!\left[\left.V^{-}\!\left(0,\cdot\right)\right|\mathscr{G}\right]<+\infty\text{ a.s.}
	\end{equation}
\end{assumption}

Finally, we impose the following growth condition on $V$.
\begin{assumption}\label{as:growthV}
	There exists a constant $\gamma>0$ and a random variable $\bar{C}\geq 0$ a.s.\ such that $\mathbb{E}_{\mathbb{P}}\!\left[\bar{C}\right]<+\infty$ and for a.e.\ $\omega$,
	\begin{equation}\label{eq:asgrowthV}
V^{+}\!\left(\lambda x,\omega\right)\leq \lambda^{\gamma}V^{+}\!\left(x,\omega\right)+\lambda^{\gamma}\bar{C}\!\left(\omega\right)%\text{, simultaneously for all }\lambda\geq1\text{ and for all }x\geq 0.
	\end{equation}
	simultaneously for all $\lambda\geq1$ and for all $x\geq 0$.
\end{assumption}

Next, we remark that denoting by $\hat{\xi}(\omega)$ the orthogonal projection of $\xi(\omega)$ on
$D(\omega)$ for some $\xi\in\Xi^d$,
we have $\hat{\xi}\in\Xi^d$ and $\langle \hat{\xi},Y\rangle=\langle \xi,Y\rangle$ a.s.,
the reader is referred to \citet[Remark~8]{carassusrasonyi2012} for further details.
This means that any portfolio can be replaced with its projection
on $D$ without changing either its value or its desirability to the investor.
We now recall that the set of all admissible strategies in $D$ is bounded.

\begin{lemma}\label{lem:rvKx}
	Assume that Assumption \ref{as:kappadelta} holds true. Given any $x_{0}\geq0$, there exists a $\mathscr{G}$-measurable, real-valued random variable $K_{x_{0}}\triangleq x_0/\beta\geq x_{0}$ such that, for every $x\in\left[0,x_{0}\right]$ and for every $\xi\in\tilde{\Xi}^{d}\!\left(x\right)$,
	we have %the inequality
	\begin{equation}
		\left\Vert\xi\right\Vert\leq K_{x_{0}} \text{ a.s.}
	\end{equation}
	%holds a.s.
\end{lemma}
\begin{proof}
	This is Lemma~2.1 in \citet{rsch06}.
\end{proof}

As for the next lemma, it will allow us to apply the Fatou lemma to a sequence of conditional expectations tending to the
essential supremum in \eqref{eq:GH} below.
\begin{lemma}\label{lem:FatouL} Assume that Assumptions \ref{as:Dlinsub}, \ref{as:kappadelta}, \ref{as:V}, \ref{as:esssupfiniteas}, \ref{as:V-} and \ref{as:growthV} hold true.
	Given any $x\geq0$, there is a non-negative random variable
	$L':~\Omega\rightarrow\mathbb{R}$ such that
	$\mathbb{E}\!\left[\left.L'\right|\mathscr{G}\right]<+\infty$ a.s., and for every
	$\xi\in\tilde{\Xi}^{d}\!\left(x\right)$ the inequality
	\begin{equation}
		V^{+}\!\left(x+\left\langle\xi\!\left(\cdot\right),Y\!\left(\cdot\right)\right\rangle,\cdot\right)\leq L_{x}(\cdot)
	\end{equation}
	holds for all $x$ with $L_{x}\triangleq\left(x^{\overline{\gamma}}+1\right)L'$, outside a fixed $\mathbb{P}$-null set.
\end{lemma}

\begin{proof}
	See Section~\ref{sec:Aux}.
\end{proof}

Now a regular version of the essential supremum is shown to exist.
\begin{lemma}\label{lem:versionesssup} Assume that Assumptions \ref{as:Dlinsub}, \ref{as:kappadelta}, \ref{as:V}, \ref{as:esssupfiniteas}, \ref{as:V-} and \ref{as:growthV} hold true.
	There exists a function $G~:~\left[\left.0,+\infty\right)\right.\times\Omega\rightarrow\mathbb{R}$ satisfying the two properties below:
	\begin{enumerate}[label=\emph{(\roman*)}]
		\item
		the function $G\!\left(x,\cdot\right)$ is a version of $\esssup_{\xi\in\Xi^{d}\!\left(x\right)}\mathbb{E}_{\mathbb{P}}\!\left[\left.V\!\left(x+\left\langle \xi\!\left(\cdot\right),Y\!\left(\cdot\right) \right\rangle,\cdot\right)\right|\mathscr{G}\right]$ for each $x\in\left[\left.0,+\infty\right)\right.$;
		
		\item
		for $\mathbb{P}$-a.e.\ $\omega\in\Omega$, the function $G\!\left(\cdot,\omega\right):~\left[\left.0,+\infty\right)\right.\rightarrow\mathbb{R}$ is non-decreasing and continuous on $\left[\left.0,+\infty\right)\right.$.%, with $G\!\left(x,\omega\right)<+\infty$ for all $x\in\left[\left.0,+\infty\right)\right.$.
	\end{enumerate}
	Furthermore, given any $\mathscr{G}$-measurable random variable $H\geq 0$ a.s.,
	\begin{equation}\label{eq:GH}
		 G\!\left(H\!\left(\cdot\right),\cdot\right)=\esssup_{\xi\in\Xi^{d}\!\left(H\right)}\mathbb{E}_{\mathbb{P}}\!\left[\left.V\!\left(H\!\left(\cdot\right)+\left\langle \xi\!\left(\cdot\right),Y\!\left(\cdot\right) \right\rangle,\cdot\right)\right|\mathscr{G}\right]\text{ a.s.}
	\end{equation}
\end{lemma}

\begin{proof}
	See Appendix~\ref{sec:Aux}.
\end{proof}

\begin{proposition}\label{prop:optonestep} Assume that Assumptions \ref{as:Dlinsub}, \ref{as:kappadelta}, \ref{as:V}, \ref{as:esssupfiniteas}, \ref{as:V-} and \ref{as:growthV} hold true.
	For every $\mathscr{G}$-measurable random variable $H\geq 0$ a.s., there exists $\widetilde{\xi}\!\left(H\right)\left(\cdot\right)\in\tilde{\Xi}^{d}\!\left(H\right)$
with
\begin{align}
		 G\!\left(H\!\left(\cdot\right),\cdot\right)%&=\esssup_{\xi\in\Xi^{d}\!\left(H\right)}\mathbb{E}_{\mathbb{P}}\!\left[\left.V\!\left(H\!\left(\cdot\right)+\left\langle \xi\!\left(\cdot\right),Y\!\left(\cdot\right) \right\rangle,\cdot\right)\right|\mathscr{G}\right]\nonumber\\
		&=\mathbb{E}_{\mathbb{P}}\!\left[\left.V\!\left(H+\left\langle \widetilde{\xi}\!\left(H\right)\left(\cdot\right),Y\!\left(\cdot\right)
		\right\rangle,\cdot\right)\right|\mathscr{G}\right]\text{ a.s.}\label{eq:xitilde}
\end{align}
\end{proposition}

\begin{proof}
	See Section \ref{sec:Aux}.
\end{proof}

%%%%%%%%%%%%%%%%%%%%%%%%%%%%%%%%%%%%%%%%%%%%%%%%%%%%%%%%%%%%%%%%%%%%%%%%%%%%%%%%%%%%%%%%%%%%%%%%%%%%%%%%%%%%%%%%%%%%%%%%%%%%%%%%%%%%%%%%%%
%%                                                                                                                                      %%
%% The multi-step case                                                                                                                  %%
%%                                                                                                                                      %%
%%%%%%%%%%%%%%%%%%%%%%%%%%%%%%%%%%%%%%%%%%%%%%%%%%%%%%%%%%%%%%%%%%%%%%%%%%%%%%%%%%%%%%%%%%%%%%%%%%%%%%%%%%%%%%%%%%%%%%%%%%%%%%%%%%%%%%%%%%

\section{The multi-step case}\label{sec:Dynamic}

In this section, we shall follow \citet{rs05,rsch06,carassusrasonyi2012}, %\citet{rsch06}, and \citet{carassusrasonyi2012}, %
and employ a dynamic programming approach to split the original optimisation problem into a number of sub-problems at different trading dates. Our goal is to invoke the results of the preceding section, thus allowing us to obtain an optimal solution at each stage. Combining them in an appropriate way will yield a globally optimal investment strategy.

%The proof our Theorem~\ref{th:optstrategy} will then consist in verifying that the conditions of the one-step case (treated in Section~\ref{sec:OneStep}) are satisfied, which will then allow us to construct an optimal portfolio. This will be accomplished in several steps, in which a dynamic programming argument will be used.% We follow the main ideas of \citet[Proposition~3.1]{rsch06}.

\begin{proof}[Proof of Theorem~\ref{th:optstrategy}]
	We must prove that some crucial assumptions of Section~\ref{sec:OneStep} are preserved at each time step. So let us start by defining
	%\begin{equation*}
	%	U_{T}\!\left(x,\omega\right)\triangleq \left\{
	%	\begin{array}{ll}
	%		u\!\left(x\right),& \text{if }x\geq 0,\\
	%		-\infty,& \text{otherwise},
	%	\end{array}
	%	\right.
	%\end{equation*}
	%for all $\omega\in\Omega$. %
	\begin{equation*}
		U_{T}\!\left(x,\omega\right)\triangleq u\!\left(x,\omega\right),\qquad x\geq0,\ \omega\in\Omega.
	\end{equation*}
	We wish to apply the results of Section~\ref{sec:OneStep} with $Y\triangleq \Delta S_{T}$, $\mathscr{G}\triangleq \mathscr{F}_{T-1}$ and $V\triangleq U_{T}$.
	%Note that $\mathscr{F}_{T}=\mathscr{F}$ by assumption.
	\begin{enumerate}[label=\emph{(\roman*)}]
		\item		
		Since Assumption~\ref{as:NA} holds by hypothesis, Theorem~3 in \citet{jacod98} implies that the affine space
		$D_{T}\!\left(\omega\right)$ is a linear subspace of $\mathbb{R}^{d}$ a.s., therefore
		Assumption~\ref{as:Dlinsub} is verified.
		It follows from Proposition~\ref{prop:quantNA} that Assumption~\ref{as:kappadelta} holds as well.
		
		\item
		We note further that Assumption~\ref{as:V} is also true from the definition of a random utility function.
%Indeed, if we fix any $x\geq 0$, then it
%		follows immediately from the definition of a random utility that the function
%		$U_{T}\!\left(x,\cdot\right):~\Omega\rightarrow\mathbb{R}$ is $\mathscr{F}_{T}$-measurable and,
%		for all $\omega$, $U_T(\omega,\cdot)$ is continuous and nondecreasing on
%		$\left[\left.0,+\infty\right)\right.$.

		\item We now claim that Assumption~\ref{as:esssupfiniteas} is satisfied. In order to show this, fix an arbitrary $x\geq 0$.
First we check that $\mathbb{E}_{\mathbb{P}}\!\left[\left.U_{T}\!\left(x+\left\langle \xi\!\left(\cdot\right),\Delta S_{T}\!\left(\cdot\right)\right\rangle,\cdot\right)\right|\mathscr{F}_{T-1}\right]$  is well-defined and finite a.s. for any given $\xi\in\Xi_{T-1}^{d}\!\left(x\right)$. It is straightforward to see that the $\mathbb{R}^{d}$-valued process defined by
		\begin{align*}
			\left(\overline{\phi}_{\xi}\right)_{t}\triangleq\left\{
				\begin{array}{ll}
					\xi,&\text{if }t=T,\\
					0,&\text{otherwise},
				\end{array}
			\right.&&
		\end{align*}
		is a portfolio in $\Psi\!\left(x\right)$, with
		\begin{align*}
			 \mathbb{E}_{\mathbb{P}}\!\left[\left.u^{+}\!\left(\Pi^{\overline{\phi}_{\xi}}_{T}\!\left(\cdot\right),\cdot\right)\right|\mathscr{F}_{T-1}\right]&=\mathbb{E}_{\mathbb{P}}\!\left[\left.u^{+}\!\left(x+\left\langle \xi\!\left(\cdot\right),\Delta S_{T}\!\left(\cdot\right)\right\rangle,\cdot\right)\right|\mathscr{F}_{T-1}\right]\\
			&=\mathbb{E}_{\mathbb{P}}\!\left[\left.U_{T}^{+}\!\left(x+\left\langle \xi\!\left(\cdot\right),\Delta S_{T}\!\left(\cdot\right)\right\rangle,\cdot\right)\right|\mathscr{F}_{T-1}\right]\text{ a.s.}
		\end{align*}
		In particular, the preceding equality and \eqref{lepra} imply that
		\begin{equation*}
			\mathbb{E}_{\mathbb{P}}\!\left[\mathbb{E}_{\mathbb{P}}\!\left[\left.U_{T}^{+}\!\left(x+\left\langle \xi\!\left(\cdot\right),\Delta S_{T}\!\left(\cdot\right)\right\rangle,\cdot\right)\right|\mathscr{F}_{T-1}\right]\right]=\mathbb{E}_{\mathbb{P}}\!\left[u^{+}\!\left(\Pi^{\overline{\phi}_{\xi}}_{T}\!\left(\cdot\right),\cdot\right)\right]<+\infty,
		\end{equation*}
		and so $\mathbb{E}_{\mathbb{P}}\!\left[\left.U_{T}^{+}\!\left(x+\left\langle \xi\!\left(\cdot\right),\Delta S_{T}\!\left(\cdot\right)\right\rangle,\cdot\right)\right|\mathscr{F}_{T-1}\right]<+\infty$ a.s. (thus, the conditional expectation is well-defined and finite a.s.\ from Remark \ref{obs:significance} \ref{obs:vstargequx0}).

Next, it can be easily shown that
		\begin{equation*}
			\left\{
			\mathbb{E}_{\mathbb{P}}\left[\left.U_{T}^{+}\!\left(x+
			\left\langle \xi\!\left(\cdot\right),\Delta S_{T}\!\left(\cdot\right)\right\rangle,
			\cdot\right)\right|\mathscr{F}_{T-1}\right]\text{: }\xi\in\Xi^{d}_{T-1}\!\left(x\right)\right\} %
		\end{equation*}
		is directed upwards, so we can find a countable sequence of random vectors $\left\{\xi_{n}\text{; }n\in\mathbb{N}\right\}\subseteq \Xi_{T-1}^{d}\!\left(x\right)$ attaining the essential supremum, i.e., such that
		\begin{multline*}
			\lim_{n\rightarrow+\infty}\mathbb{E}_{\mathbb{P}}\left[\left.U_{T}^{+}\!\left(x+\left\langle \xi_{n}\!\left(\cdot\right),\Delta S_{T}\!\left(\cdot\right)\right\rangle,\cdot\right)\right|\mathscr{F}_{T-1}\right]\\
			=\esssup_{\xi\in\Xi^{d}_{T-1}\!\left(x\right)}\mathbb{E}_{\mathbb{P}}\left[\left.U_{T}^{+}\!\left(x+\left\langle \xi\!\left(\cdot\right),\Delta S_{T}\!\left(\cdot\right)\right\rangle,\cdot\right)\right|\mathscr{F}_{T-1}\right]\text{ a.s.}
		\end{multline*}
		in a non-decreasing way. Therefore, it follows from the Monotone Convergence Theorem and from the definition of conditional expectations that
		\begin{align*}
			 \lefteqn{\mathbb{E}_{\mathbb{P}}\!\left[\esssup_{\xi\in\Xi^{d}_{T-1}\!\left(x\right)}\mathbb{E}_{\mathbb{P}}\left[\left.U_{T}^{+}\!\left(x+\left\langle \xi\!\left(\cdot\right),\Delta S_{T}\!\left(\cdot\right)\right\rangle,\cdot\right)\right|\mathscr{F}_{T-1}\right]\right]}\\
			 &\hspace{4cm}=\lim_{n\rightarrow+\infty}\mathbb{E}_{\mathbb{P}}\!\left[\mathbb{E}_{\mathbb{P}}\left[\left.U_{T}^{+}\!\left(x+\left\langle \xi_{n}\!\left(\cdot\right),\Delta S_{T}\!\left(\cdot\right)\right\rangle,\cdot\right)\right|\mathscr{F}_{T-1}\right]\right]\\
			&\hspace{4cm}=\sup_{n\in\mathbb{N}}\mathbb{E}_{\mathbb{P}}\left[U_{T}^{+}\!\left(x+\left\langle \xi_{n}\!\left(\cdot\right),\Delta S_{T}\!\left(\cdot\right)\right\rangle,\cdot\right)\right] \\
&\hspace{4cm}=\sup_{n\in\mathbb{N}}\mathbb{E}_{\mathbb{P}}\!\left[u^{+}\!\left(\Pi^{\overline{\phi}_{n}}_{T}\!\left(\cdot\right),\cdot\right)\right]<+\infty,
		\end{align*}
where $\overline{\phi}_{n}\triangleq \overline{\phi}_{\xi_{n}}$ and we invoked \eqref{lepra} once again.
		%\begin{multline*}
%			 \mathbb{E}_{\mathbb{P}}\!\left[\esssup_{\xi\in\Xi^{d}_{T-1}\!\left(x\right)}\mathbb{E}_{\mathbb{P}}\left[\left.U_{T}^{+}\!\left(x+\left\langle \xi\!\left(\cdot\right),\Delta S_{T}\!\left(\cdot\right)\right\rangle,\cdot\right)\right|\mathscr{F}_{T-1}\right]\right]\\
%			 =\sup_{n\in\mathbb{N}}\mathbb{E}_{\mathbb{P}}\!\left[u^{+}\!\left(\Pi^{\overline{\phi}_{n}}_{T}\!\left(\cdot\right),\cdot\right)\right]<+\infty,
%		\end{multline*}
		
Hence $\esssup_{\xi\in\Xi^{d}_{T-1}\!\left(x\right)}\mathbb{E}_{\mathbb{P}}\left[\left.U_{T}^{+}\!\left(x+\left\langle \xi\!\left(\cdot\right),\Delta S_{T}\!\left(\cdot\right)\right\rangle,\cdot\right)\right|\mathscr{F}_{T-1}\right]<+\infty$ a.s.%, as we claimed.
		
		\item
		The next step is to show that Assumption~\ref{as:V-} holds as well. In fact, due to Assumption~\ref{as:u0}, it is immediate that
		\begin{equation*}
			 \mathbb{E}_{\mathbb{P}}\!\left[\mathbb{E}_{\mathbb{P}}\!\left[\left.U_{T}^{-}\!\left(0,\cdot\right)\right|\mathscr{F}_{T-1}\right]\right]=\mathbb{E}_{\mathbb{P}}\!\left[U_{T}^{-}\!\left(0,\cdot\right)\right]=\mathbb{E}_{\mathbb{P}}\!\left[u^{-}\!\left(0,\cdot\right)\right]<+\infty,
		\end{equation*}
		so $\mathbb{E}_{\mathbb{P}}\!\left[\left.U_{T}^{-}\!\left(0,\cdot\right)\right|\mathscr{F}_{T-1}\right]<+\infty$ a.s.
				
		\item
		Lastly, let the constant $\overline{\gamma}>0$ and the integrable random variable
		$C(\cdot) \geq 0$ a.s. be those given by Assumption~\ref{as:growthu} and Lemma~\ref{lem:growthu}. Then, for a.e.\ $\omega\in\Omega$,
		\begin{equation}\label{eq:UTlambda}
			U_{T}^{+}\!\left(\lambda x,\omega\right)=u^{+}\!\left(\lambda x,\omega\right)\leq \lambda^{\overline{\gamma}}u^{+}\!\left(x,\omega\right)+C(\omega)\lambda^{\overline{\gamma}}=
\lambda^{\overline{\gamma}}U_{T}^{+}\!\left(x,\omega\right)+C(\omega)\lambda^{\overline{\gamma}}
		\end{equation}
		simultaneously for all $\lambda\geq 1$ and $x\geq 0$.
	\end{enumerate}
	
	Now, by Lemma~\ref{lem:versionesssup}, there exists a function %
	%\begin{equation*}
	$	G_{T-1}:~\left[\left.0,+\infty\right)\right.\times\Omega\rightarrow\mathbb{R}$ %
	%\end{equation*}
	such that, for every $\omega$ in a $\mathbb{P}$-full measure set, the function $G_{T-1}\!\left(\cdot,\omega\right):~\left[\left.0,+\infty\right)\right.\rightarrow\mathbb{R}$ is non-decreasing and continuous on $\left[\left.0,+\infty\right)\right.$. Moreover, for every $x\in\left[\left.0,+\infty\right)\right.$,
	\begin{equation*}
		 G_{T-1}\!\left(x,\cdot\right)=\esssup_{\xi\in\Xi_{T-1}^{d}\!\left(x\right)}\mathbb{E}_{\mathbb{P}}\!\left[\left.U_{T}\!\left(x+\left\langle \xi\!\left(\cdot\right),\Delta S_{T}\!\left(\cdot\right) \right\rangle,\cdot\right)\right|\mathscr{F}_{T-1}\right]\text{ a.s.}
	\end{equation*}
	In addition, Proposition~\ref{prop:optonestep} gives us, for each $H\in\Xi_{T-1}^1$, an $\mathscr{F}_{T-1}$-measurable function
	$\widetilde{\xi}_{T}\left(H\right)(\cdot)~:~\Omega\rightarrow\mathbb{R}^{d}$ such that
	\begin{eqnarray}\label{tomtom}
G_{T-1}\left(H(\cdot),\cdot\right) &=&		\esssup_{\xi\in\Xi_{T-1}^{d}\!\left(H\right)}\mathbb{E}_{\mathbb{P}}\!\left[\left.U_{T}\!\left(
		H(\cdot)+\left\langle \xi\!\left(\cdot\right),\Delta S_{T}\!\left(\cdot\right) \right\rangle,\cdot\right)\right|\mathscr{F}_{T-1}\right]\\
\nonumber		&=& \mathbb{E}_{\mathbb{P}}\!\left[\left.U_{T}\!\left(H(\cdot)+\left\langle \widetilde{\xi}_{T}
		\left(H\right)(\cdot),\Delta S_{T}\!\left(\cdot\right) \right\rangle,\cdot\right)\right|\mathscr{F}_{T-1}\right]\text{ a.s.}
	\end{eqnarray}
	
	Let us now proceed to the next stage of dynamic programming. Let $U_{T-1}:~\left[\left.0,+\infty\right)\right.\times\Omega\rightarrow\mathbb{R}$ be the function given by %
	%\begin{equation*}
	%	U_{T-1}\!\left(x,\omega\right)\triangleq \left\{
	%	\begin{array}{ll}
	%		G_{T-1}\!\left(x,\omega\right),& \text{if }x\geq 0,\\
	%		-\infty,& \text{otherwise},
	%	\end{array}
	%	\right.
	%\end{equation*}
	$U_{T-1}\!\left(x,\omega\right)\triangleq G_{T-1}\!\left(x,\omega\right)$. %
	As before, we would like to use the results of Section~\ref{sec:OneStep}, this time with $Y\triangleq \Delta S_{T-1}$, $\mathscr{G}\triangleq \mathscr{F}_{T-2}$ and $V\triangleq U_{T-1}$.
	\begin{enumerate}[label=\emph{(\roman*')}]
		\item
		Assumptions~\ref{as:Dlinsub} and~\ref{as:kappadelta} are both true, as before.
		
		\item
		Next, we prove that Assumption~\ref{as:V} holds. In fact, given any $x\geq 0$, the function $U_{T-1}\!\left(x,\cdot\right):~\Omega\rightarrow\mathbb{R}$ %equals $U_{T-1}\!\left(x,\omega\right)=G_{T-1}\!\left(x,\omega\right)$ for every $\omega$, hence it %
		is $\mathscr{F}_{T-1}$-measurable. On the other hand, for a.e. $\omega$, we have by definition of $U_{T-1}$ that $U_{T-1}\!\left(\cdot,\omega\right)$ is a non-decreasing continuous function on $\left[\left.0,+\infty\right)\right.$.%, and $U_{T-1}\!\left(x,\omega\right)=G_{T-1}\!\left(x,\omega\right)<+\infty$ for all $x\geq 0$.
		
		\item
		We show that we also have Assumption~\ref{as:esssupfiniteas}. Indeed, letting $x\geq 0$ be arbitrary, but fixed, it can be easily checked, in the same way as before (the construction of the portfolio becoming more involved, but totally analogous), that for every $\xi\in\Xi_{T-2}^{d}\!\left(x\right)$, the conditional expectation
		\begin{equation*}
			\mathbb{E}_{\mathbb{P}}\left[\left.U_{T-1}\!\left(x+\left\langle \xi\!\left(\cdot\right),\Delta S_{T-1}\!\left(\cdot\right)\right\rangle,\cdot\right)\right|\mathscr{F}_{T-2}\right]
		\end{equation*}
		is not only well-defined, but also finite a.s. Furthermore,
		\begin{equation*}
			\esssup_{\xi\in\Xi^{d}_{T-2}\!\left(x\right)}\mathbb{E}_{\mathbb{P}}\left[\left.U_{T-1}^{+}\!\left(x+\left\langle \xi\!\left(\cdot\right),\Delta S_{T-1}\!\left(\cdot\right)\right\rangle,\cdot\right)\right|\mathscr{F}_{T-2}\right]<+\infty\text{ a.s.,}
		\end{equation*}
		as desired.
		
		\item
		We proceed with the proof that Assumption~\ref{as:V-} is also verified. Given any $x\geq 0$, it is clear that
		\begin{equation*}
			U_{T-1}\!\left(x,\cdot\right)=G_{T-1}\!\left(x,\cdot\right)\geq \mathbb{E}_{\mathbb{P}}\left[\left.U_{T}\!\left(x,\cdot\right)\right|\mathscr{F}_{T-1}\right]\text{ a.s.,}%=\mathbb{E}_{\mathbb{P}}\left[\left.u\!\left(x,\cdot\right)\right|\mathscr{F}_{T-1}\right]\text{ a.s.,}
		\end{equation*}
		where the inequality is due to $0\in\Xi^{d}_{T-1}\!\left(x\right)$ and to the definition of
		the essential supremum. Now we can use Jensen's inequality (for the conditional expectation) to obtain %
		\begin{align*}
			 \mathbb{E}_{\mathbb{P}}\!\left[\mathbb{E}_{\mathbb{P}}\!\left[\left.U_{T-1}^{-}\!\left(0,\cdot\right)\right|\mathscr{F}_{T-2}\right]\right]&=\mathbb{E}_{\mathbb{P}}\!\left[U_{T-1}^{-}\!\left(0,\cdot\right)\right]\\
			 &\leq\mathbb{E}_{\mathbb{P}}\!\left[\mathbb{E}_{\mathbb{P}}\!\left[\left.U_{T}^{-}\!\left(0,\cdot\right)\right|\mathscr{F}_{T-1}\right]\right]=\mathbb{E}_{\mathbb{P}}\!\left[U_{T}^{-}\!\left(0,\cdot\right)\right]=\mathbb{E}_{\mathbb{P}}\!\left[u^{-}\!\left(0,\cdot\right)\right],
		\end{align*}
		which in turn implies (recall Assumption~\ref{as:u0}) that  $\mathbb{E}_{\mathbb{P}}\!\left[\left.U_{T-1}^{-}\!\left(0,\cdot\right)\right|\mathscr{F}_{T-2}\right]<+\infty$ a.s. Note in passing that we have also  proved that
\begin{eqnarray}
\label{negativer}
\mathbb{E}_{\mathbb{P}}\!\left[U_{T-1}^{-}\!\left(0,\cdot\right)\right] \leq \mathbb{E}_{\mathbb{P}}\!\left[u^{-}\!\left(0,\cdot\right)\right].
\end{eqnarray}
		
		\item\label{item:3.3.v'}
		Take the constant $\overline{\gamma}>0$ and the integrable random variable $C(\cdot) \geq 0$ a.s. to be as in Assumption~\ref{as:growthu} and Lemma~\ref{lem:growthu}.
We have that, for every $\lambda\geq 1$ and $x\geq 0$,
		\begin{align*}
			U_{T-1}^{+}\!\left(\lambda x,\cdot\right)&\leq\mathbb{E}_{\mathbb{P}}\!\left[\left.U_{T}^{+}\!\left(\lambda x+\left\langle \widetilde{\xi}_{T}\!\left(\lambda x\right)\left(\cdot\right),\Delta S_{T}\!\left(\cdot\right) \right\rangle,\cdot\right)\right|\mathscr{F}_{T-1}\right]\\
&\leq \lambda^{\overline{\gamma}}\mathbb{E}_{\mathbb{P}}\!\left[\left.U^+_{T}\!\left(x+\left\langle
\widetilde{\xi}_{T}\!\left(\lambda x\right)\left(\cdot\right)
/\lambda,\Delta S_{T}\!\left(\cdot\right) \right\rangle,\cdot\right)\right|\mathscr{F}_{T-1}\right]+\lambda^{\overline{\gamma}}\mathbb{E}_{\mathbb{P}}\!\left[\left.C(\cdot)\right|\mathscr{F}_{T-1}\right]\text{ a.s.,}
		\end{align*}
		where the first inequality follows from the conditional Jensen inequality and from \eqref{tomtom} with $H=\lambda x$, the second one uses \eqref{eq:UTlambda}. Moreover, it is easy to see that $\widetilde{\xi}_{T}\!\left(\lambda x\right)\left(\cdot\right)/\lambda\in\Xi^{d}_{T-1}\!\left(x\right)$ and we obtain
	\begin{align*}
		\lefteqn{\mathbb{E}_{\mathbb{P}}\!\left[\left.U^+_{T}\!\left(x+\left\langle
\widetilde{\xi}_{T}\!\left(\lambda x\right)\left(\cdot\right)
/\lambda,\Delta S_{T}\!\left(\cdot\right) \right\rangle,\cdot\right)\right|\mathscr{F}_{T-1}\right]}\\
		&=\mathbb{E}_{\mathbb{P}}\!\left[\left.U_{T}\!\left(x+\left\langle
\widetilde{\xi}_{T}\!\left(\lambda x\right)\left(\cdot\right)
/\lambda,\Delta S_{T}\!\left(\cdot\right) \right\rangle,\cdot\right)\right|\mathscr{F}_{T-1}\right]\\
		&\qquad+ \mathbb{E}_{\mathbb{P}}\!\left[\left.U^-_{T}\!\left(x+\left\langle
\widetilde{\xi}_{T}\!\left(\lambda x\right)\left(\cdot\right)
/\lambda,\Delta S_{T}\!\left(\cdot\right) \right\rangle,\cdot\right)\right|\mathscr{F}_{T-1}\right]\\
		&\leq\mathbb{E}_{\mathbb{P}}\!\left[\left.U_{T}\!\left(x+\left\langle
\widetilde{\xi}_{T}\!\left(\lambda x\right)\left(\cdot\right)
/\lambda,\Delta S_{T}\!\left(\cdot\right) \right\rangle,\cdot\right)\right|\mathscr{F}_{T-1}\right]+ \mathbb{E}_{\mathbb{P}}\!\left[\left.U^-_{T}\!\left(0,\cdot\right)\right|\mathscr{F}_{T-1}\right]\\
& \leq U_{T-1}\!\left(x,\cdot\right)+ \mathbb{E}_{\mathbb{P}}\!\left[\left.U^-_{T}\!\left(0,\cdot\right)\right|\mathscr{F}_{T-1}\right]\text{ a.s.}
	\end{align*}
	So we conclude that, for every $\lambda\geq1$ and $x\geq0$, %
	\begin{equation*}
		U_{T-1}^{+}\!\left(\lambda x,\cdot\right)\leq\lambda^{\overline{\gamma}}U_{T-1}^{+}\!\left(x,\cdot\right)+\lambda^{\overline{\gamma}}\bar{C}(\cdot) \text{ a.s.}
	\end{equation*}
	for $\bar{C}(\cdot)\triangleq\mathbb{E}_{\mathbb{P}}\!\left[\left.C(\cdot)\right|\mathscr{F}_{T-1}\right]+ \mathbb{E}_{\mathbb{P}}\!\left[\left.U^-_{T}\!\left(0,\cdot\right)\right|\mathscr{F}_{T-1}\right]<+\infty$
a.s.\ by Assumption \ref{as:V-} for $V=U_{T}$. Moreover, $\mathbb{E}_{\mathbb{P}}\!\left[\bar{C}(\cdot)\right]=\mathbb{E}_{\mathbb{P}}\!\left[C(\cdot)\right]+\mathbb{E}_{\mathbb{P}}\!\left[u^{-}\!\left(0,\cdot\right)\right]<+\infty$ by Assumption~\ref{as:u0}. %
%\footnote{Note for the next step that \eqref{negativer} will allow to conclude that $\mathbb{E}_{\mathbb{P}}[U_{T-1}^-\!\left(0,\cdot\right)]<\infty$}
		Using the regularity of the paths of $U_{T-1}$, we get that, for a.e.\ $\omega$ the inequality %
		%\begin{equation*}
		$	U_{T-1}^{+}\!\left(\lambda x,\omega\right)\leq\lambda^{\overline{\gamma}}U_{T-1}^{+}\!\left(x,\omega\right)+\lambda^{\overline{\gamma}}\bar{C}(\omega)$ %
		%\end{equation*}
		holds simultaneously for all $\lambda\geq 1$ and $x\geq 0$, so Assumption~\ref{as:growthV} is verified.
	\end{enumerate}
	
	Consequently, we can apply Lemma~\ref{lem:versionesssup} and Proposition~\ref{prop:optonestep} to
	obtain functions $G_{T-2}$ and $\widetilde{\xi}_{T-1}$ satisfying the desired properties.
	Proceeding in a similar way for the remaining values of $t\in\left\{T-2,\ldots,1\right\}$
	(noting that, for the next step, \eqref{negativer} will allow us to conclude in step \ref{item:3.3.v'} that
	$\mathbb{E}_{\mathbb{P}}\!\left[U_{T-1}^-\!\left(0,\cdot\right)\right]<+\infty$), we construct the functions $U_{T-2},\ldots,U_{1},U_{0}$ and $\widetilde{\xi}_{T-2},\ldots,\widetilde{\xi}_{1}$.
	
We then inductively define $\Pi^{\phi^*}_{0}=x_{0}$, $\phi_{1}^{*}\!\left(\cdot\right)\triangleq\tilde{\xi}_{1}\!\left(x_{0}\right)\left(\cdot\right)$ and $\phi_{t}^{*}\!\left(\cdot\right)\triangleq\tilde{\xi}_{t}\!\left(\Pi^{\phi^*}_{t-1}\right)\left(\cdot\right)$. The remainder of the proof, showing optimality of $\phi^{*}$, unfolds exactly as the proof of Proposition~3.2 in \citet{rsch06}. We report it here for convenience of the reader. Joint measurability of $\widetilde{\xi}_{t}$ ensures that $\phi^{*}$ is a predictable process with respect to the given filtration. Recall that we have proved that
\begin{equation*}
	\esssup_{\xi\in\Xi^{d}_{t-1}\!\left(x\right)}\mathbb{E}_{\mathbb{P}}\left[\left.U_{t}^{+}\!\left(x+\left\langle \xi\!\left(\cdot\right),\Delta S_{t}\!\left(\cdot\right)\right\rangle,\cdot\right)\right|\mathscr{F}_{t-1}\right]<+\infty\text{ a.s.,}
\end{equation*}
so the all the conditional expectations below exist and are finite.

It is easy to see that $\phi_{t}^{*}\!\left(\cdot\right)\in\Xi_{t-1}\!\left(\Pi^{\phi^*}_{t-1}\right)$, and Proposition \ref{prop:optonestep} shows that, for $t\in\left\{1,\ldots,T\right\}$,
\begin{align*}
	 \mathbb{E}_{\mathbb{P}}\!\left[\left.U_{t}\!\left(\Pi^{\phi^{*}}_{t}\!\left(\cdot\right),\cdot\right)\right|\mathscr{F}_{t-1}\right]&=
\mathbb{E}_{\mathbb{P}}\!\left[\left.U_{t}\!\left(\Pi^{\phi^{*}}_{t-1}\!\left(\cdot\right)+\left\langle
\tilde{\xi}_{t}\!\left(\Pi^{\phi^{*}}_{t-1}\right)\!\left(\cdot\right),\Delta S_{t}\!\left(\cdot\right)\right\rangle,\cdot\right)\right|\mathscr{F}_{t-1}\right]\\
	&=U_{t-1}\!\left(\Pi^{\phi^{*}}_{t-1}\!\left(\cdot\right),\cdot\right)\text{ a.s.}
\end{align*}
So $\mathbb{E}_{\mathbb{P}}\!\left[\left.U_{t}\!\left(\Pi^{\phi^{*}}_{t}\!\left(\cdot\right),\cdot\right)\right|\mathscr{F}_{0}\right]=\mathbb{E}_{\mathbb{P}}\!\left[\left.U_{t-1}\!\left(\Pi^{\phi^{*}}_{t-1}\!\left(\cdot\right),\cdot\right)\right|\mathscr{F}_{0}\right]$ a.s., and we deduce that
\begin{equation}\label{bintou}
	 \mathbb{E}_{\mathbb{P}}\!\left[\left.U_{T}\!\left(\Pi^{\phi^{*}}_{T}\!\left(\cdot\right),\cdot\right)\right|\mathscr{F}_{0}\right]=\mathbb{E}_{\mathbb{P}}\!\left[\left.U_{1}\!\left(\Pi^{\phi^{*}}_{1}\!\left(\cdot\right),\cdot\right)\right|\mathscr{F}_{0}\right]=U_{0}\!\left(x_{0}\right)\text{ a.s.}
\end{equation}

Now let $\phi \in \Psi\!\left(x_{0}\right)$. Clearly, $\phi_{T}\in \Xi_{T-1}\!\left(\Pi_{T-1}^{\phi}\right)$. Using Proposition \ref{prop:optonestep} again, we obtain
\begin{equation*}
	\mathbb{E}_{\mathbb{P}}\!\left[\left.U_{T}\!\left(\Pi^{\phi}_{T-1}\!\left(\cdot\right)+\left\langle\phi_{T}\!\left(\cdot\right),\Delta S_{T}\!\left(\cdot\right)\right\rangle,\cdot\right)\right|\mathscr{F}_{T-1}\right]\leq U_{T-1}\!\left(\Pi_{T-1}\!\left(\cdot\right),\cdot\right)\text{ a.s.}
\end{equation*}
Iterating the same argument we get
\begin{eqnarray}
\label{polution}
\mathbb{E}_{\mathbb{P}}\!\left[\left.U_{T}\!\left(\Pi^{\phi}_{T-1}\!\left(\cdot\right)+\left\langle\phi_{T}\!\left(\cdot\right),\Delta S_{T}\!\left(\cdot\right)\right\rangle,\cdot\right)\right|\mathscr{F}_{0}\right]\leq U_{0}\!\left(x_{0}\right) \mbox{ a.s.}
\end{eqnarray}
and the theorem is proved, recalling \eqref{bintou}.
\end{proof}

\begin{proof}[Proof of Theorem~\ref{th:StinW}]

The proof unfolds exactly as that of Theorem~\ref{th:optstrategy}, the only
difference residing, for each time stage $t\in \{T,\ldots,1\}$ , in the verification that Assumption \ref{as:esssupfiniteas} is valid for the function $U_t$  $:$ $[0, \infty) \times \Omega \to \mathbb{R}$ which is defined recursively as in the proof of Theorem~\ref{th:optstrategy}. More precisely \emph{(i)}, \emph{(ii)}, \emph{(iv)}, \emph{(v)}, \emph{(i')}, \emph{(ii')}, \emph{(iv')} and \emph{(v')} of the proof of   Theorem~\ref{th:optstrategy} are exactly the same and will  prove \emph{(iii)} and \emph{(iii')} (and Assumption \ref{as:esssupfiniteas}) using \eqref{eclipse} below. Note that the assumption that $v^*(x_0)<\infty$ in the proof of Theorem~\ref{th:optstrategy} was only used in \emph{(iii)} and \emph{(iii')}.

We prove by backward induction on $t$ that there exists some $J_{t}\in\mathscr{W}$ such that
\begin{eqnarray}
\label{eclipse}
U_{t}\!\left(x,\omega\right)\leq J_{t}\!\left(\omega\right)\left[x^{\overline{\gamma}}+1\right]
\end{eqnarray}
simultaneously for all $x \geq 0$, for $\omega$ outside a $\mathbb{P}$-null set.
Starting with $t=T$, we set $U_{T}\!\left(x,\cdot\right)\triangleq u\!\left(x,\cdot\right)$.
Assumption~\ref{as:growthu} shows that, outside a $\mathbb{P}$-null set, for all $x\geq\overline{x}$,
	\begin{equation*}
		U_{T}\!\left(x,\cdot\right)=u\!\left(x,\cdot\right)\leq (x/\overline{x})^{\overline{\gamma}}[u^{+}\!\left(\overline{x},\cdot\right)+c\!\left(\cdot\right)],
	\end{equation*}
	and for $0\leq x<\overline{x}$,
	\begin{equation*}
		U_{T}\!\left(x,\cdot\right)\leq U^{+}_{T}\!\left(\overline{x},\cdot\right)=u^{+}\!\left(\overline{x},\cdot\right),
	\end{equation*}
	so we can set
	\begin{equation*}
		 J_{T}\!\left(\cdot\right)\triangleq\max\!\left\{[u^{+}(\overline{x},\cdot)+c\!\left(\cdot\right)]/\overline{x}^{\overline{\gamma}},u^+({\overline{x}},\cdot)\right\},
	\end{equation*}
	and the latter is clearly in $\mathscr{W}$ by our assumptions.

Now, let us assume that the statement has been shown for $T,T-1,\ldots,t+1$. Using Lemma~\ref{lem:rvKx}, we can estimate
	\begin{align*}
	 U_{t}\!\left(x,\cdot\right)&\triangleq \esssup_{\xi\in\widetilde{\Xi}_{t}^{d}\!\left(x\right)}\mathbb{E}_{\mathbb{P}}\!\left[\left.U_{t+1}\!\left(x+\left\langle\xi\!\left(\cdot\right),\Delta S_{t+1}\!\left(\cdot\right)\right\rangle,\cdot\right)\right|\mathscr{F}_{t}\right]\\
		&\leq \esssup_{\xi\in\widetilde{\Xi}_{t-1}^{d}\!\left(x\right)}\mathbb{E}_{\mathbb{P}}\!\left[\left.J_{t+1}\!\left(\cdot\right)\left[x+\left\Vert\xi\!\left(\cdot\right)\right\Vert\,\left\Vert\Delta S_{t+1}\!\left(\cdot\right)\right\Vert\right]^{\overline{\gamma}}+J_{t+1}\!\left(\cdot\right)\right|\mathscr{F}_{t}\right]\\
		&\leq x^{\overline{\gamma}}\mathbb{E}_{\mathbb{P}}\!\left[\left.J_{t+1}\!\left(\cdot\right)\left[1+\left\Vert\Delta S_{t+1}\!\left(\cdot\right)\right\Vert/\beta_{t+1}\!\left(\cdot\right)\right]^{\overline{\gamma}}\right|\mathscr{F}_{t}\right]+\mathbb{E}_{\mathbb{P}}\!\left[\left.J_{t+1}\!\left(\cdot\right)\right|\mathscr{F}_{t}\right]\text{ a.s.,}
	\end{align*}
	so we may set $J_{t}\!\left(\cdot\right)\triangleq\mathbb{E}_{\mathbb{P}}\!\left[\left.J_{t+1}\!\left(\cdot\right)\left[1+\left\Vert\Delta S_{t+1}\!\left(\cdot\right)\right\Vert/\beta_{t+1}\!\left(\cdot\right)\right]^{\overline{\gamma}}\right|\mathscr{F}_{t}\right]\in\mathscr{W}$. This inequality has been obtained for each $x \geq 0$ a.s., but using the regularity of the paths of $U_{t}$ we get it for all $x \geq 0$ on a common $\mathbb{P}$-full measure set.

Now we show how \eqref{eclipse} implies that Assumption \ref{as:esssupfiniteas} holds true and thus replaces \emph{(iii)} and \emph{(iii')} in the proof of
Theorem~\ref{th:optstrategy}. The argument is the same as above using that $J_{t+1} \geq 0$ a.s. and thus
$U^+_{t}\!\left(x,\omega\right)\leq J_{t}\!\left(\omega\right)\left[x^{\overline{\gamma}}+1\right]$ for all $x \geq 0$ on a common $\mathbb{P}$-full measure set. Thus
\begin{align*}
			& \esssup_{\xi\in\widetilde{\Xi}_{t-1}^{d}\!\left(x\right)}\mathbb{E}_{\mathbb{P}}\!\left[\left.U^+_{t}\!\left(x+\left\langle\xi\!\left(\cdot\right),\Delta S_{t}\!\left(\cdot\right)\right\rangle,\cdot\right)\right|\mathscr{F}_{t-1}\right]\\
		&\leq \esssup_{\xi\in\widetilde{\Xi}_{t}^{d}\!\left(x\right)}\mathbb{E}_{\mathbb{P}}\!\left[\left.J_{t}\!\left(\cdot\right)\left[x+\left\Vert\xi\!\left(\cdot\right)\right\Vert\,\left\Vert\Delta S_{t}\!\left(\cdot\right)\right\Vert\right]^{\overline{\gamma}}+J_{t}\!\left(\cdot\right)\right|\mathscr{F}_{t-1}\right]\\
		&\leq (x^{\overline{\gamma}}+1) J_{t-1}\!\left(\cdot\right)<+\infty\text{ a.s.}
	\end{align*}
and Assumption \ref{as:esssupfiniteas} holds true. The end of proof of Theorem~\ref{th:StinW} follows verbatim the one of Theorem~\ref{th:optstrategy}.

Note that here we obtain that $v^{*}(x_0)<\infty$ for all $x_0\geq 0$. Indeed from \eqref{polution} where $\phi \in \Psi(x_0)$ is arbitrary, we get that $v^{*}\!\left(x_0\right)\leq \mathbb{E}_{\mathbb{P}}\!\left[U_{0}\!\left(x_0,\cdot\right)\right]$.  Thus
$v^{*}\!\left(x_0\right)\leq (1+x_0^{\overline{\gamma}})\mathbb{E}_{\mathbb{P}}\!\left[J_{0}\!\left(\cdot\right)\right]<+\infty$.

\end{proof}

%%%%%%%%%%%%%%%%%%%%%%%%%%%%%%%%%%%%%%%%%%%%%%%%%%%%%%%%%%%%%%%%%%%%%%%%%%%%%%%%%%%%%%%%%%%%%%%%%%%%%%%%%%%%%%%%%%%%%%%%%%%%%%%%%%%%%%%%%%
%%                                                                                                                                      %%
%% Appendix: proofs and auxiliary results                                                                                               %%
%%                                                                                                                                      %%
%%%%%%%%%%%%%%%%%%%%%%%%%%%%%%%%%%%%%%%%%%%%%%%%%%%%%%%%%%%%%%%%%%%%%%%%%%%%%%%%%%%%%%%%%%%%%%%%%%%%%%%%%%%%%%%%%%%%%%%%%%%%%%%%%%%%%%%%%%

%\appendix
\section{Appendix}\label{sec:Aux}

%Except where explicitly stated otherwise,
We are staying in the setting of Section~\ref{sec:OneStep}.

\begin{example}\label{ex:boundedAEinf}
	It is not difficult to find (non-random) utilities which are bounded above and yet have non-zero (actually, infinite)
	asymptotic elasticity, as the example below shows.
	The construction below is inspired by the proof of Lemma~6.5 in \citet{kramkov99}. Let $f:~\left[\left.0,+\infty\right)\right.\rightarrow\mathbb{R}$ be the function which takes the values
	
	\begin{align*}
	%\begin{equation*}
		&f\!\left(n\right)\triangleq \frac{1}{2}-\frac{1}{n+1}=\frac{n-1}{2\left(n+1\right)},\\
	%\end{equation*}
	%\begin{equation*}
		&f\!\left(n+1/2-a_{n}\right)\triangleq f\!\left(n\right)+a_{n},\\
	%\end{equation*}
	%\begin{equation*}
		&f\!\left(n+1/2+a_{n}\right)\triangleq f\!\left(n+1\right)-a_{n},
	%\end{equation*}
	\end{align*}
		%\begin{align*}
		%	f\!\left(n\right)&\triangleq \frac{1}{2}-\frac{1}{n+1}=\frac{n-1}{2\left(n+1\right)},\\
		%	f\!\left(n+a_{n}\right)&\triangleq \frac{n-1}{2\left(n+1\right)}-\frac{a_{n}}{2\left(n+1\right)\left(n+2\right)},\\
		%	f\!\left(n+1-a_{n}\right)&\triangleq \frac{n}{2\left(n+2\right)}-\frac{a_{n}}{2\left(n+2\right)\left(n+3\right)},\\
		%\end{align*}
		with %
		%\begin{equation*}
		%	a_{n}\triangleq \frac{n^{3}+6n^{2}+10n+3}{2n^{3}+12n^{2}+21n+10}\in\left(0,\frac{1}{2}\right),
		%\end{equation*}
		$a_{n}\triangleq 1/\left[4\left(n+1\right)\left(n+2\right)\right]$ for every $n\in\mathbb{N}$, and which is linear between the points where it has been defined.
		
		Clearly $f\!\left(+\infty\right)=1/2$ and $f\!\left(1\right)=0$. We also note that $f\!\left(0\right)=-1/2>-\infty$. %, and $\lim_{n\rightarrow+\infty}a_{n}=1/2$ in an increasing way. %
		Moreover, the piecewise linearity of $f$ and trivial computations yield
		%\begin{align*}
			 %u'\!\left(x\right)&=\frac{1}{1-2a_{n}}\left(\frac{n}{2\left(n+2\right)}-\frac{a_{n}}{2\left(n+2\right)\left(n+3\right)}-\frac{n-1}{2\left(n+1\right)}-\frac{a_{n}}{2\left(n+1\right)\left(n+2\right)}\right)\\
		%	&=1
		%\end{align*}
		\begin{equation*}
			f'\!\left(x\right)=\frac{f\!\left(n+1/2+a_{n}\right)-f\!\left(n+1/2-a_{n}\right)}{2a_{n}}=1
		\end{equation*}
		for any $x\in\left(n+1/2-a_{n},n+1/2+a_{n}\right)$, so in particular $f'\!\left(n+1/2\right)$ equals $1$. Furthermore, we have the following inequality,% for the average utility at $n+1/2$,
		\begin{equation*}
			\frac{f\!\left(n+1/2\right)}{n+1/2}\leq \frac{f\!\left(n+1\right)}{n+1/2}=\frac{n}{\left(n+2\right)\left(2n+1\right)},
		\end{equation*}
		thus combining all of the above gives %
		%\begin{equation*}
		%	\frac{\left(n+1/2\right)f'\!\left(n+1/2\right)}{f\!\left(n+1/2\right)} \xrightarrow[n\rightarrow +\infty]{}+\infty,
		%\end{equation*}
		$\lim_{n\rightarrow +\infty}\left(n+1/2\right)f'\!\left(n+1/2\right)/f\!\left(n+1/2\right)=+\infty$, %
		and hence $AE_{+}\!\left(u\right)=+\infty$ in the classical sense of \citet{kramkov99}. We finish by noticing that, as in the proof of Lemma~6.5 in \citet{kramkov99}, $f$ can be slightly modified in such a way that it becomes smooth and our conclusion is still valid. So, as mentioned in Remark \ref{obs:growthu},
Lemma~6.3 of \citet{kramkov99} applies and
	\begin{equation*}
		\limsup_{x\rightarrow+\infty}\frac{x\,f'\!\left(x\right)}{f\!\left(x\right)}=
		\inf\!\left\{\gamma>0\text{ : }\exists\,\overline{x}\geq0 \text{ s.t.\ a.e.\ }
		f\!\left(\lambda x,\cdot\right)\leq\lambda^{\gamma}f\!\left(x,\cdot\right),
		\ \forall\,\lambda\geq1,\ \forall\,x\geq\overline{x}\right\},
	\end{equation*}
	with the usual convention that the infimum of the empty set is $+\infty$, and $f$ has an infinite asymptotic elasticity in both senses. Nevertheless, choosing $\overline{x}=1$, we have that $0 \leq f(x) \leq 1/2$ for $x \geq \overline{x}$, and Assumption~\ref{as:growthu} holds true for $u(x)=f(x)$.
\end{example}

%=========================================================================================================================================

\begin{proof}[Proof of Lemma~\ref{lem:growthu}]\label{proof:lem:growthu}
	%Set $\widehat{x}\triangleq\max\!\left\{1,\overline{x}\right\}$, where $\overline{x}$ is
	%that of Assumption~\ref{as:growthu}, and
	%Let us begin by noticing that the inequality is trivial on the set $\left\{\omega\in\Omega\text{: }u\!\left(\lambda x,\omega\right)<0,\; \forall \lambda \geq 1, \; \forall x \geq 0 \right\}$. So if this set is of full measure, the proof is complete. Else we  will work on $E\triangleq\left\{\omega\in\Omega\text{: }u\!\left(\lambda x,\omega\right)\geq 0,\; \forall \lambda \geq 1, \; \forall x \geq 0\right\}$, where of course $u\!\left(\lambda x,\omega\right)=u^{+}\!\left(\lambda x,\omega\right)$.
%	
	Consider an arbitrary $x\in [0,\overline{x})$.
	Then we can use the fact that $u$ is non-decreasing %
	and inequality \eqref{eq:asgrowthu} to obtain for a.e. $\omega \in \Omega$ that %
	\begin{equation*}
		u\!\left(\lambda x,\omega\right)\leq u\!\left(\lambda \overline{x},\omega\right)\leq \lambda^{\overline{\gamma}}u\!\left(\overline{x},\omega\right)+\lambda^{\overline{\gamma}}c\!\left(\omega\right)\leq \lambda^{\overline{\gamma}}u^{+}\!\left(\overline{x},\omega\right)+\lambda^{\overline{\gamma}}c\!\left(\omega\right) %
	\end{equation*}
	for any $\lambda\geq 1$. As $c \geq 0$ a.s. this implies that
$u^{+}\!\left(\lambda x,\omega\right) \leq \lambda^{\overline{\gamma}}u^{+}\!\left(\overline{x},\omega\right)+\lambda^{\overline{\gamma}}c\!\left(\omega\right).$
	On the other hand, for  a.e. $\omega \in \Omega$ and every $x\geq\overline{x}$, we have by \eqref{eq:asgrowthu} that %
	\begin{equation*}
		u\!\left(\lambda x,\omega\right)\leq \lambda^{\overline{\gamma}}\left[u\!\left(x,\omega\right)+c\!\left(\omega\right)\right]\leq \lambda^{\overline{\gamma}}\left[u^{+}\!\left(x,\omega\right)+c\!\left(\omega\right)\right] %
	\end{equation*}
	for all $\lambda\geq 1$. Using again that $c \geq 0$ a.s. we get that
	$u^{+}\!\left(\lambda x,\omega\right)\leq \lambda^{\overline{\gamma}}\left[u^{+}\!\left(x,\omega\right)+c\!\left(\omega\right)\right].$
	Hence, choosing $C\!\left(\omega\right)\triangleq u^+\!\left(\overline{x},\omega\right)+c\!\left(\omega\right)\geq0$ and combining the  previous inequalities we get for a.e. $\omega \in \Omega$
	\begin{align*}
		u^{+}\!\left(\lambda x,\omega\right)&\leq\max\left\{\lambda^{\overline{\gamma}}\left[u^{+}\!\left(\overline{x},\omega\right)+c\!\left(\omega\right)\right],\lambda^{\overline{\gamma}}\left[u^{+}\!\left(x,\omega\right)+c\!\left(\omega\right)\right]\right\}%\\
		\leq% \lambda^{\overline{\gamma}}\left[u^{+}\!\left(x,\omega\right)+u^+\!\left(\overline{x},\omega\right)+c\!\left(\omega\right)\right]=%
		\lambda^{\overline{\gamma}}u^{+}\!\left(x,\omega\right)+\lambda^{\overline{\gamma}}C\!\left(\omega\right)
	\end{align*}
	for all $\lambda\geq 1$, $x\geq 0$. Lastly, note that $\mathbb{E}_{\mathbb{P}}\!\left[C\right]<+\infty$ since $\mathbb{E}_{\mathbb{P}}\!\left[u^+\!\left(\overline{x},\cdot\right)\right]<+\infty$ and $\mathbb{E}_{\mathbb{P}}\!\left[c\right]<+\infty$ by Assumption~\ref{as:growthu}.
\end{proof}

%=========================================================================================================================================

\begin{proof}[Proof of Lemma~\ref{lem:FatouL}]

	Let $\Theta$ denote the set of functions from $\{1,\ldots,d\}$ to $\{-\sqrt{d},\sqrt{d}\}$, and let $x>0$. Then we have by Lemma~\ref{lem:rvKx} that, for all $\xi\in\widetilde{\Xi}^d(x)$,
	\begin{equation*}
		x+\left\langle\xi,Y\right\rangle\leq x\left(1+\left\Vert Y\right\Vert/\beta\right)\leq x\left[1+\left(1/\beta\right)\max_{\tau\in\Theta}\left\langle\tau,Y\right\rangle\right]\text{ a.s.},
	\end{equation*}
	hence
	\begin{align*}
		V^{+}\!\left(x+\left\langle\xi\!\left(\cdot\right),Y\!\left(\cdot\right)\right\rangle,\cdot\right)&\leq V^{+}\!\left(x\left[1+\left(1/\beta\!\left(\cdot\right)\right)\max_{\tau\in\Theta}\left\langle\tau,Y\!\left(\cdot\right)\right\rangle\right],\cdot\right)\\
		&\leq \sum_{\tau\in\Theta}V^{+}\!\left(x\left[1+\left\langle\tau/\beta\!\left(\cdot\right),Y\!\left(\cdot\right)\right\rangle\right],\cdot\right)\text{ a.s.,}
	\end{align*}
	where we define $V^{+}\!\left(x,\omega\right)\triangleq0$ for $x<0$.

	We know from Proposition~4.2 of \citet{rsch06} that there is a random set $M\!\left(1\right)\in\mathscr{G}\otimes\mathscr{B}\!\left(\mathbb{R}^d\right)$ such that $\xi\in M\!\left(1\right)$ a.s.\ if and only if $\xi\in\tilde{\Xi}^{d}\!\left(1\right)$. Denoting the linear span of $M\!\left(1\right)\!\left(\omega\right)$ by $R\!\left(1\right)\!\left(\omega\right)$, $R\!\left(1\right)$ is again in $\mathscr{G}\otimes\mathscr{B}\!\left(\mathbb{R}^d\right)$ by Proposition~4.3 of \citet{rsch06}. It suffices to prove our lemma separately on the sets $\left\{\omega\in\Omega\text{ : }\text{dim}\,R\!\left(1\right)\!\left(\omega\right)=k\right\}$, with $k\in\left\{0,\ldots,d\right\}$, since these sets are in $\mathscr{G}$ (see the proof of Proposition~4.3 in \citet{rsch06}). Applying verbatim the arguments of Lemma~2.3 in \citet{rsch06}, %
%\textcolor[rgb]{0.98,0.00,0.00}{je sais que ma vesrion était fausse avec le relative interior mais peut-on être un peu plus précis? de plus si $\varepsilon_{\tau}$ est une variable aléatoire ce n'est pas clair pour moi que l'on peut appliquer l'AE. }
one can show the existence of $g\in\Xi^{d}\!\left(1\right)$ and of $\varepsilon_{\tau}\in \Xi^{1}$ with $\varepsilon_{\tau}\in (0,1)$ such that $\widetilde{g}_{\tau}\triangleq g+\varepsilon_{\tau}\left({\tau}/\beta-g\right)$ belongs to $M(1)$ and thus to $\Xi^{d}\!\left(1\right)$ (with the notation of the cited lemma, $\mathscr{G}=\mathscr{H}$, $g=\rho$, $K\theta_i=\tau/\beta$, $\varepsilon_{\tau}=\psi_i$ and $i=\tau$). It follows that, for $x\leq 1$,
	\begin{equation*}
		V^{+}\!\left(x+\left\langle\xi\!\left(\cdot\right),Y\!\left(\cdot\right)\right\rangle,\cdot\right)\leq \sum_{\tau\in\Theta}V^{+}\!\left(1+\left\langle\tau/\beta\!\left(\cdot\right),Y\!\left(\cdot\right)\right\rangle,\cdot\right)\text{ a.s.},
	\end{equation*}
	and almost surely for $x>1$,
	\begin{equation*}
		V^{+}\!\left(x+\left\langle\xi\!\left(\cdot\right),Y\!\left(\cdot\right)\right\rangle,\cdot\right)\leq \sum_{\tau\in\Theta}x^{\overline{\gamma}}\left[V^{+}\!\left(1+\left\langle\tau/\beta\!\left(\cdot\right),Y\!\left(\cdot\right)\right\rangle,\cdot\right)+
\overline{C}\!\left(\cdot\right)\right].
	\end{equation*}
	by Assumption~\ref{as:growthV}. Note that if $\tau$ is such that $\{1+\left\langle\tau/\beta\!\left(\cdot\right),Y\!\left(\cdot\right)\right\rangle < 0\}$ has strictly positive probability then on this set the preceding inequality holds true since $\overline{C} \geq 0$ a.s.

Applying the same assumption again (and with the same remark), we get a.s. that
	\begin{align*}
		\lefteqn{V^{+}\left(1+\left\langle\frac{{\tau}}{\beta\!\left(\cdot\right)},Y\!\left(\cdot\right)\right\rangle,\cdot\right)}\\
		 &\qquad\qquad\qquad\leq\frac{1}{{\varepsilon_{\tau}\!\left(\cdot\right)}^{\overline{\gamma}}}\left[V^{+}\!\left(\varepsilon_{\tau}\!\left(\cdot\right)\left[1+\left\langle\frac{{\tau}}{\beta\!\left(\cdot\right)},Y\!\left(\cdot\right)\right\rangle\right],\cdot\right)+\overline{C}\!\left(\cdot\right)\right]\\
		 &\qquad\qquad\qquad\leq\frac{1}{{\varepsilon_{\tau}\!\left(\cdot\right)}^{\overline{\gamma}}}\left[V^{+}\!\left(\varepsilon_{\tau}\!\left(\cdot\right)\left[1+\left\langle g\!\left(\cdot\right),Y\!\left(\cdot\right)\right\rangle\right]+\varepsilon_{\tau}\!\left(\cdot\right)\left\langle\frac{\tau}{\beta\!\left(\cdot\right)}-g\!\left(\cdot\right),Y\!\left(\cdot\right)\right\rangle,\cdot\right)+\overline{C}\!\left(\cdot\right)\right]\\
  	&\qquad\qquad\qquad\leq\frac{1}{{\varepsilon_{\tau}\!\left(\cdot\right)}^{\overline{\gamma}}}\left[V^{+}\!\left(1+\left\langle g\!\left(\cdot\right),Y\!\left(\cdot\right)\right\rangle+\varepsilon_{\tau}\!\left(\cdot\right)\left\langle\frac{\tau}{\beta\!\left(\cdot\right)}-g\!\left(\cdot\right),Y\!\left(\cdot\right)\right\rangle,\cdot\right)+\overline{C}\!\left(\cdot\right)\right]\\
 	 &\qquad\qquad\qquad=\frac{1}{{\varepsilon_{\tau}\!\left(\cdot\right)}^{\overline{\gamma}}}\left[V^{+}\!\left(1+\left\langle\widetilde{g}_{\tau}\!\left(\cdot\right),Y\!\left(\cdot\right)\right\rangle,\cdot\right)+\overline{C}\!\left(\cdot\right)\right],
\end{align*}
where the last inequality holds true since $1+\left\langle g,Y\right\rangle\geq0$ a.s. Now let
\begin{equation*}
	L'\!\left(\cdot\right)\triangleq\sum_{\tau \in \Theta}\left(\frac{1}{{\varepsilon_{\tau}\!\left(\cdot\right)}^{\overline{\gamma}}}\left[V^{+}\!\left(1+\left\langle\widetilde{g}_{\tau}\!\left(\cdot\right),Y\!\left(\cdot\right)\right\rangle,\cdot\right)+\overline{C}\!\left(\cdot\right)\right]+\overline{C}\!\left(\cdot\right)\right)+V^{+}\!\left(0,\cdot\right),
\end{equation*}
where the last term is added to cover the case $x=0$ as well. By assumption, $\mathbb{E}_{\mathbb{P}}\!\left[\left.\overline{C}\right|\mathscr{G}\right]<+\infty$ a.s., and by Assumption~\ref{as:esssupfiniteas} we have
$\mathbb{E}_{\mathbb{P}}\!\left[\left.V^{+}\!\left(1+\left\langle\widetilde{g}_{\tau}\!\left(\cdot\right),Y\!\left(\cdot\right)\right\rangle,
\cdot\right)\right|\mathscr{G}\right]<+\infty$ and $\mathbb{E}_{\mathbb{P}}\!\left[\left.V^{+}\!\left(0,
\cdot\right)\right|\mathscr{G}\right]<+\infty$ a.s., so the proof is complete since $\varepsilon_{\tau}$ is $\mathscr{G}$-measurable.
\end{proof}

%=========================================================================================================================================

\begin{proof}[Proof of Lemma~\ref{lem:versionesssup}]\label{proof:versionesssup}
	Let us first choose, for each positive rational number $q$, a version $F\!\left(q,\omega\right)$ of $\esssup_{\xi\in\Xi^{d}\!\left(q\right)}\mathbb{E}_{\mathbb{P}}\!\left[\left.V\!\left(q+\left\langle \xi\!\left(\cdot\right),Y\!\left(\cdot\right) \right\rangle,\cdot\right)\right|\mathscr{G}\right]$.
	
Next, for any pair $q_{1}<q_{2}$ of positive rational numbers, and any $\xi\in\Xi^{d}\!\left(q_{1}\right)$, one has $\xi\in\Xi^{d}\!\left(q_{2}\right)$, as obviously $q_{1}+\left\langle \xi,Y\right\rangle<q_{2}+\left\langle \xi,Y\right\rangle$. So we have by Assumption~\ref{as:V} that
	\begin{equation*}
		V\!\left(q_{1}+\left\langle \xi\!\left(\omega\right),Y\!\left(\omega\right)\right\rangle,\omega\right)\leq V\!\left(q_{2}+\left\langle \xi\!\left(\omega\right),Y\!\left(\omega\right)\right\rangle,\omega\right)
	\end{equation*}
	for $\mathbb{P}$-a.e.\ $\omega\in\Omega$. We then conclude from the monotonicity of the conditional expectation and taking essential supremum that %
	%\begin{equation*}
%		\mathbb{E}_{\mathbb{P}}\!\left[\left.V\!\left(q_{1}+\left\langle \xi\!\left(\cdot\right),Y\!\left(\cdot\right)\right\rangle,\cdot\right)\right|\mathscr{G}\right]\leq\mathbb{E}_{\mathbb{P}}\!\left[\left.V\!\left(q_{2}+\left\langle \xi\!\left(\cdot\right),Y\!\left(\cdot\right)\right\rangle,\cdot\right)\right|\mathscr{G}\right]\leq F\!\left(q_{2},\cdot\right)\text{ a.s.,}
%	\end{equation*}
%	and it follows from the definition of essential supremum, combined with the arbitrarity of $\xi$, that
$F\!\left(q_{1},\cdot\right)\leq F\!\left(q_{2},\cdot\right)$ a.s. Similarly, for every positive $q\in\mathbb{Q}$, $F\!\left(q,\omega\right)<+\infty$ from Assumption~\ref{as:esssupfiniteas}. Thus one can find a set $A \in \mathscr{G}$ of full probability such that, for every $\omega\in A$, the mapping $q \mapsto F\!\left(q,\omega\right)$ is non-decreasing and finite-valued on the set of positive rational numbers.

	So let us specify, for each $x\in\left[\left.0,+\infty\right)\right.$ and $\omega\in A$,
	\begin{equation*}
		G\!\left(x,\omega\right)\triangleq\inf_{\substack{q\in\mathbb{Q}\\q> x}} F\!\left(q,\omega\right),
	\end{equation*}
	and define $G\!\left(x,\omega\right)=0$ if $\omega \in \Omega \setminus A$. Clearly, when $x\geq 0$ is in $\mathbb{Q}$ then $G\!\left(x,\cdot\right)\geq F\!\left(x,\cdot\right)$ a.s. In addition, for each $x\geq 0$ we have that $G\!\left(x,\cdot\right)$ is $\mathscr{G}$-measurable. We shall split the remainder of the proof into five separate parts.
	\begin{enumerate}[label=\emph{(\roman*)}]
		\item
		With the above definition, it is straightforward to check that, for each $\omega\in\Omega$, the function $G\!\left(\cdot,\omega\right)$ is non-decreasing. It is also clear that $G\!\left(x,\omega\right)<+\infty$ for all $x\geq 0$ and $\omega \in \Omega$.
		
		\item\label{partversion}
		We proceed to show that, for all $x\in\left[\left.0,+\infty\right)\right.$,
		\begin{equation*}
			G\!\left(x,\cdot\right)=\esssup_{\xi\in\Xi^{d}\!\left(x\right)}\mathbb{E}_{\mathbb{P}}\!\left[\left.V\!\left(x+\left\langle \xi\!\left(\cdot\right),Y\!\left(\cdot\right) \right\rangle,\cdot\right)\right|\mathscr{G}\right] \text{ a.s.}
		\end{equation*}
		In order to do so, let us fix an arbitrary $x\in\left[\left.0,+\infty\right)\right.$. Then,
		for every $q\in\mathbb{Q}$, $q>x$, the inequality
		\begin{equation*}
			\esssup_{\xi\in\Xi^{d}\!\left(x\right)}\mathbb{E}_{\mathbb{P}}\!\left[\left.V\!\left(x+\left\langle \xi\!\left(\cdot\right),Y\!\left(\cdot\right) \right\rangle,\cdot\right)\right|\mathscr{G}\right]\leq F\!\left(q,\cdot\right)
		\end{equation*}
		holds a.s., thus we get
		\begin{equation*}
			\esssup_{\xi\in\Xi^{d}\!\left(x\right)}\mathbb{E}_{\mathbb{P}}\!\left[\left.V\!\left(x+\left\langle \xi\!\left(\cdot\right),Y\!\left(\cdot\right) \right\rangle,\cdot\right)\right|\mathscr{G}\right]\leq G\!\left(x,\cdot\right)\text{ a.s.}
		\end{equation*}
		%for $\mathbb{P}$-a.e.\ $\omega\in\Omega\setminus N$.
		
		It remains to verify that the reverse inequality is also true (except possibly on a set of measure zero). This will be achieved in three steps.

		\begin{enumerate}[label=\emph{(\alph*)}]
		
		\item		
		Let us start by taking a strictly decreasing sequence $\left\{q_{n}\text{; }n\in\mathbb{N}\right\}$
		of rational numbers satisfying $x<q_{n}<x+1$ and $\lim_{n\rightarrow+\infty} q_{n}=x$.
		Now, given any $n\in\mathbb{N}$, it is straightforward that the family %
		%\begin{equation*}
		$	\left\{\mathbb{E}_{\mathbb{P}}\left[\left.V\!\left(q_{n}+\left\langle \xi\!\left(\cdot\right),Y\!\left(\cdot\right) \right\rangle,\cdot\right)\right|\mathscr{G}\right]\text{; }\xi\in\Xi^{d}\!\left(q_{n}\right)\right\}$ %
		%\end{equation*}
		is directed upwards, %(see \Cref{lem:dirdown}), %
		therefore one can find  $\zeta_n\in\Xi^d(q_n)$ such that
		\begin{equation*}
			\mathbb{E}_{\mathbb{P}}\left[\left.V\!\left(q_{n}+\left\langle \zeta_{n}\!\left(\cdot\right),Y\!\left(\cdot\right) \right\rangle,\cdot\right)\right|\mathscr{G}\right]\geq F\!\left(q_{n},\cdot\right)-\frac{1}{n}\text{ a.s.}
		\end{equation*}
		
		\item
		Next, fix an arbitrary $n$. It was observed above that $\zeta_{n}\in\Xi^{d}\!\left(q_{n}\right)\subseteq\Xi^{d}\!\left(x+1\right)$. Thus, taking $\widehat{\zeta}_{n}$ to be its projection on $D$, we know %from \Cref{lem:EVxiequal} and the preceding equations %
		that
		\begin{align}
			\mathbb{E}_{\mathbb{P}}\left[\left.V\!\left(q_{n}+\left\langle \widehat{\zeta}_{n}\!\left(\cdot\right),Y\!\left(\cdot\right) \right\rangle,\cdot\right)\right|\mathscr{G}\right]&=\mathbb{E}_{\mathbb{P}}\left[\left.V\!\left(q_{n}+\left\langle \zeta_{n}\!\left(\cdot\right),Y\!\left(\cdot\right) \right\rangle,\cdot\right)\right|\mathscr{G}\right]\nonumber\\
			&\geq F\!\left(q_{n},\cdot\right)-\frac{1}{n}\text{ a.s.}\label{eq:esssupaux1}
		\end{align}
		Moreover, Lemma~\ref{lem:rvKx} allows us to conclude that $\lVert\widehat{\zeta}_{n}\rVert\leq K_{x+1}$ a.s. Therefore,
		we can extract a random subsequence $\left\{\widehat{\zeta}_{n_{k}}\text{; }k\in\mathbb{N}\right\}$ such that %
		%\begin{equation*}
		$	\lim_{k\rightarrow+\infty} \widehat{\zeta}_{n_{k}}=\zeta$ % \text{ a.s.,}
		%\end{equation*}
		a.s., %
		for some $\mathscr{G}$-measurable random variable $\zeta$ (see Lemma 2 of \cite{kabanov01}). 
		But then
		\begin{equation*}
			x+\left\langle \zeta\!\left(\omega\right),Y\!\left(\omega\right) \right\rangle=\lim_{k\rightarrow+\infty} \left(q_{n_{k}\!\left(\omega\right)}+\left\langle \widehat{\zeta}_{n_{k}\!\left(\omega\right)}\!\left(\omega\right),Y\!\left(\omega\right) \right\rangle\right)\geq 0
		\end{equation*}
		for $\mathbb{P}$-a.e.\ $\omega\in\Omega$, i.e.\ $\zeta\in\Xi^{d}\!\left(x\right)$, which in turn implies that
		\begin{equation}\label{eq:esssupaux2}
			\esssup_{\xi\in\Xi^{d}\!\left(x\right)}\mathbb{E}_{\mathbb{P}}\!\left[\left.V\!\left(x+\left\langle \xi\!\left(\cdot\right),Y\!\left(\cdot\right) \right\rangle,\cdot\right)\right|\mathscr{G}\right]\geq \mathbb{E}_{\mathbb{P}}\!\left[\left.V\!\left(x+\left\langle \zeta\!\left(\cdot\right),Y\!\left(\cdot\right) \right\rangle,\cdot\right)\right|\mathscr{G}\right]\text{a.s.}
		\end{equation}
		
		\item
		Finally, let us define the random variables $f_{k}:~\Omega\rightarrow\mathbb{R}$ as follows,
		\begin{equation*}
			f_{k}\!\left(\omega\right)\triangleq V\!\left(q_{n_{k}\!\left(\omega\right)}+\left\langle \widehat{\zeta}_{n_{k}\!\left(\omega\right)}\!\left(\omega\right),Y\!\left(\omega\right) \right\rangle,\omega\right),\qquad \omega\in\Omega.
		\end{equation*}
		By construction of the sequence $\left\{q_{n}\text{; }n\in\mathbb{N}\right\}$ and of the random subsequence $\left\{\widehat{\zeta}_{n_{k}}\text{; }k\in\mathbb{N}\right\}$, and by the continuity of the paths of $V$ (see Assumption~\ref{as:V}), it is clear that %
		%\begin{equation*}
		$	\lim_{k\rightarrow+\infty} f_{k}=V\!\left(x+\left\langle\zeta\!\left(\cdot\right),Y\!\left(\cdot\right) \right\rangle,\cdot\right)$ a.s. %\text{ a.s.}
		%\end{equation*}
		We further observe that, for $\mathbb{P}$-a.e.\ $\omega\in\Omega$,
		\begin{equation*}
			f_{k}\!\left(\omega\right)\leq
			V\!\left(x+1+\left\langle \widehat{\zeta}_{n_{k}\!\left(\omega\right)}\!\left(\omega\right),Y\!\left(\omega\right) \right\rangle,\omega\right)\leq\left[\left(1+x\right)^{\overline{\gamma}}+1\right]L'\!\left(\omega\right),
		\end{equation*}
		where the first inequality follows from the monotonicity of $V$ (again we refer to Assumption~\ref{as:V}), and the second inequality is a simple consequence of Lemma~\ref{lem:FatouL} combined with the fact that
		\begin{equation*}
			x+1+\left\langle \widehat{\zeta}_{n_{k}\!\left(\cdot\right)}\!\left(\cdot\right),Y\!\left(\cdot\right) \right\rangle\geq q_{n_{k}\!\left(\cdot\right)}+\left\langle \widehat{\zeta}_{n_{k}\!\left(\cdot\right)}\!\left(\cdot\right),Y\!\left(\cdot\right) \right\rangle\geq 0\text{ a.s.}
		\end{equation*}
		Hence, we may apply Fatou's lemma (for the limit superior) to conclude that
		\begin{align}
			\mathbb{E}_{\mathbb{P}}\!\left[\left.V\!\left(x+\left\langle \zeta\!\left(\cdot\right),Y\!\left(\cdot\right) \right\rangle,\cdot\right)\right|\mathscr{G}\right]&\geq \limsup_{k\rightarrow+\infty} \mathbb{E}_{\mathbb{P}}\!\left[\left.f_{k}\right|\mathscr{G}\right]\nonumber\\
			&\geq\liminf_{k\rightarrow+\infty} \left[F\!\left(q_{n_{k}\!\left(\cdot\right)},\cdot\right)-\frac{1}{n_k(\cdot)}\right]\geq \inf_{n\in\mathbb{N}} F\!\left(q_{n},\cdot\right)\text{ a.s.,}\label{eq:esssupaux3}
		\end{align}
		from \eqref{eq:esssupaux1} applied on $\{n_k=i\}$ for $i \geq k$.
		
		Combining equations \eqref{eq:esssupaux2} and \eqref{eq:esssupaux3} finally gives the intended inequality %
		\begin{equation*}
			\esssup_{\xi\in\Xi^{d}\!\left(x\right)}\mathbb{E}_{\mathbb{P}}\!\left[\left.V\!\left(x+\left\langle \xi\!\left(\cdot\right),Y\!\left(\cdot\right) \right\rangle,\cdot\right)\right|\mathscr{G}\right]\geq \inf_{n\in\mathbb{N}} F\!\left(q_{n},\cdot\right)\geq G\!\left(x,\cdot\right) \text{ a.s.}
		\end{equation*}
		%$	\esssup_{\xi\in\Xi^{d}\!\left(x\right)}\mathbb{E}_{\mathbb{P}}\!\left[\left.V\!\left(x+\left\langle \xi\!\left(\cdot\right),Y\!\left(\cdot\right) \right\rangle,\cdot\right)\right|\mathscr{G}\right]\geq \inf_{n\in\mathbb{N}} F\!\left(q_{n},\cdot\right)\geq G\!\left(x,\cdot\right)$ a.s.
		\end{enumerate}
	
		\item
		Thirdly, $G$ is, by construction, right-continuous. Moreover it is easy to see that
$$G(x,\omega)=1_A(\omega) \inf_{n \in \mathbb{N}} \{F(r_n,\omega)\bbOne_{\left\{r_n >x\right\}} +(+\infty) \bbOne_{\left\{r_n \leq x\right\}}\}$$
where $\{r_n; \, n \in \mathbb{N}\}$ is an enumeration of $\mathbb{Q}$ (we make the usual convention that $0 \times \infty = 0$). Recalling that
$\omega \to F(r_n,\omega)$ is $\mathscr{G}$-measurable for each $r_n$, it follows that $G$ is measurable with
respect to the product $\sigma$-algebra $\mathscr{B}\!\left(\left[\left.0,+\infty\right)\right.\right)\otimes\mathscr{G}$.

		\item\label{item:Hcountablestep}
		Now consider an arbitrary $\mathscr{G}$-measurable random variable $H\geq 0$ a.s. We wish to show that
		\begin{equation}\label{pluie}
	 G\!\left(H\!\left(\cdot\right),\cdot\right)=\esssup_{\xi\in\Xi^{d}\!\left(H\right)}\mathbb{E}_{\mathbb{P}}\!\left[\left.V\!\left(H\!\left(\cdot\right)+\left\langle \xi\!\left(\cdot\right),Y\!\left(\cdot\right) \right\rangle,\cdot\right)\right|\mathscr{G}\right]\mbox{ a.s.}
		\end{equation}
				
		We have just proved above that \eqref{pluie} holds true for a non-negative constant $H$. It is easy to show that it holds also true for  $\mathscr{G}$-measurable countable step-functions $H$.
			
		Next, suppose $H$ is any bounded, $\mathscr{G}$-measurable, non-negative (a.s.) random variable, so there exists some constant $M>0$ such that $H\leq M$ a.s. It is a well-known fact that we can take a non-increasing sequence $\left\{H_{n}\text{; }n\in\mathbb{N}\right\}$ of $\mathscr{G}$-measurable step-functions converging to $H$ a.s., and such that, for every $n\in\mathbb{N}$, $H_{n}\leq M$ a.s. Then, fixing an arbitrary $\xi\in\Xi^{d}\!\left(H\right)$, we have for every $n\in\mathbb{N}$ that %
		%\begin{equation*}
		$	H_{n}+\left\langle \xi,Y\right\rangle\geq H+\left\langle \xi,Y\right\rangle\geq 0$ a.s., %\text{ a.s.,}
		%\end{equation*}
		therefore
		\begin{align*}
			 G\!\left(H_{n}\!\left(\cdot\right),\cdot\right)&=\esssup_{\zeta\in\Xi^{d}\!\left(H_{n}\right)}\mathbb{E}_{\mathbb{P}}\!\left[\left.V\!\left(H_{n}\!\left(\cdot\right)+\left\langle \zeta\!\left(\cdot\right),Y\!\left(\cdot\right) \right\rangle,\cdot\right)\right|\mathscr{G}\right]\\
			&\geq \mathbb{E}_{\mathbb{P}}\!\left[\left.V\!\left(H_{n}\!\left(\cdot\right)+\left\langle \xi\!\left(\cdot\right),Y\!\left(\cdot\right) \right\rangle,\cdot\right)\right|\mathscr{G}\right]\text{ a.s.}
		\end{align*}
		(recall that \eqref{pluie} is true for step-functions), which in turn yields
		\begin{equation*}
			\liminf_{n\rightarrow+\infty}G\!\left(H_{n}\!\left(\cdot\right),\cdot\right)\geq\liminf_{n\rightarrow+\infty} \mathbb{E}_{\mathbb{P}}\!\left[\left.V\!\left(H_{n}\!\left(\cdot\right)+\left\langle \xi\!\left(\cdot\right),Y\!\left(\cdot\right) \right\rangle,\cdot\right)\right|\mathscr{G}\right]\text{ a.s.}
		\end{equation*}
		On the one hand, we get by right-continuity of $G$ that %
		%\begin{equation*}
		$	\lim_{n\rightarrow+\infty}G\!\left(H_{n}\!\left(\cdot\right),\cdot\right)=G\!\left(H\!\left(\cdot\right),\cdot\right)$ a.s. %\text{ a.s.}
		%\end{equation*}
		On the other hand, we can apply Fatou's lemma (for the limit inferior, see Assumption~\ref{as:V-}) to conclude% that
		%\begin{align*}
			%\lefteqn{\liminf_{n\rightarrow+\infty} \mathbb{E}_{\mathbb{P}}\!\left[\left.V\!\left(H_{n}\!\left(\cdot\right)+\left\langle \xi\!\left(\cdot\right),Y\!\left(\cdot\right) \right\rangle,\cdot\right)\right|\mathscr{G}\right]}\\
			%&\geq \liminf_{n\rightarrow+\infty} \mathbb{E}_{\mathbb{P}}\!\left[\left.V^{+}\!\left(H_{n}\!\left(\cdot\right)+\left\langle \xi\!\left(\cdot\right),Y\!\left(\cdot\right) \right\rangle,\cdot\right)\right|\mathscr{G}\right]-\limsup_{n\rightarrow+\infty} \mathbb{E}_{\mathbb{P}}\!\left[\left.V^{-}\!\left(H_{n}\!\left(\cdot\right)+\left\langle \xi\!\left(\cdot\right),Y\!\left(\cdot\right) \right\rangle,\cdot\right)\right|\mathscr{G}\right]\\
		%	&\geq \mathbb{E}_{\mathbb{P}}\!\left[\left.V\!\left(H\!\left(\cdot\right)+\left\langle \xi\!\left(\cdot\right),Y\!\left(\cdot\right) \right\rangle,\cdot\right)\right|\mathscr{G}\right],
		%\end{align*}
		\begin{multline*}
			\liminf_{n\rightarrow+\infty} \mathbb{E}_{\mathbb{P}}\!\left[\left.V\!\left(H_{n}\!\left(\cdot\right)+\left\langle \xi\!\left(\cdot\right),Y\!\left(\cdot\right) \right\rangle,\cdot\right)\right|\mathscr{G}\right]
			\geq \mathbb{E}_{\mathbb{P}}\!\left[\left.V\!\left(H\!\left(\cdot\right)+\left\langle \xi\!\left(\cdot\right),Y\!\left(\cdot\right) \right\rangle,\cdot\right)\right|\mathscr{G}\right]\text{ a.s.},
		\end{multline*}
		hence %
		%\begin{equation*}
		$	\esssup_{\xi\in\Xi^{d}\!\left(H\right)}\mathbb{E}_{\mathbb{P}}\!\left[\left.V\!\left(H\!\left(\cdot\right)+\left\langle \xi\!\left(\cdot\right),Y\!\left(\cdot\right) \right\rangle,\cdot\right)\right|\mathscr{G}\right]\leq G\!\left(H\!\left(\cdot\right),\cdot\right)$ a.s.\ (by the arbitrariness of $\xi\in\Xi^{d}\!\left(H\right)$). %\text{ a.s.}
		%\end{equation*}
		Now, to prove the reverse inequality, we can construct (as in part \ref{partversion} of this proof) a sequence $\left\{\zeta_{n}\text{; }n\in\mathbb{N}\right\}$ such that, for every $n\in\mathbb{N}$, we have $\zeta_{n}\in\Xi^{d}\!\left(H_{n}\right)$, $\zeta_{n}\!\left(\omega\right)\in D\!\left(\omega\right)$ for $\mathbb{P}$-a.e.\ $\omega\in\Omega$, and
		\begin{align}
			G\!\left(H_{n}\!\left(\cdot\right),\cdot\right) -\frac{1}{n}&=\esssup_{\xi\in\Xi^{d}\!\left(H_{n}\right)}\mathbb{E}_{\mathbb{P}}\!\left[\left.V\!\left(H_{n}\!\left(\cdot\right)+\left\langle \xi\!\left(\cdot\right),Y\!\left(\cdot\right) \right\rangle,\cdot\right)\right|\mathscr{G}\right]-\frac{1}{n}\nonumber\\
			&\leq\mathbb{E}_{\mathbb{P}}\!\left[\left.V\!\left(H_{n}\!\left(\cdot\right)+\left\langle \zeta_{n}\!\left(\cdot\right),Y\!\left(\cdot\right) \right\rangle,\cdot\right)\right|\mathscr{G}\right]\text{ a.s.}\label{neige}
		\end{align}
		We remark further that each $\zeta_{n}$ belongs to $\Xi^{d}\!\left(M\right)$ (because $M+\left\langle \zeta_{n},Y\right\rangle\geq H_{n}+\left\langle \zeta_{n},Y\right\rangle\geq 0$ a.s.), so by Lemma~\ref{lem:rvKx} there exists a random variable $K_{M}$ such that $\left\Vert\zeta_{n}\right\Vert\leq K_{M}$ a.s. Therefore we can select a random subsequence $\left\{\zeta_{n_{k}}\text{; }k\in\mathbb{N}\right\}$ with %
		%\begin{equation*}
		$	\lim_{k\rightarrow+\infty}\zeta_{n_{k}}=\zeta$ a.s., %\text{ a.s.,}
		%\end{equation*}
		for some $\mathscr{G}$-measurable $\zeta$. Clearly, %
		\begin{equation*}
			H+\left\langle \zeta,Y\right\rangle=\lim_{k\rightarrow+\infty}\left(H_{n_{k}}+\left\langle \zeta_{n_{k}},Y\right\rangle\right) \text{ a.s.},
		\end{equation*}
		and for every $k\in\mathbb{N}$,
		\begin{equation*}
			H_{n_{k}}+\left\langle \zeta_{n_{k}},Y\right\rangle=\sum_{i=k}^{+\infty}\left(H_{i}+\left\langle \zeta_{i},Y\right\rangle\right)\bbOne_{\left\{n_{k}\!\left(\cdot\right)=i\right\}}\geq0\text{ a.s.},
		\end{equation*}
		hence $\zeta\in\Xi^{d}\!\left(H\right)$. Consequently,
		\begin{multline*}
			\esssup_{\xi\in\Xi^{d}\!\left(H\right)}\mathbb{E}_{\mathbb{P}}\!\left[\left.V\!\left(H\!\left(\cdot\right)+\left\langle \xi\!\left(\cdot\right),Y\!\left(\cdot\right) \right\rangle,\cdot\right)\right|\mathscr{G}\right]
			\geq\mathbb{E}_{\mathbb{P}}\!\left[\left.V\!\left(H\!\left(\cdot\right)+\left\langle \zeta\!\left(\cdot\right),Y\!\left(\cdot\right) \right\rangle,\cdot\right)\right|\mathscr{G}\right]\text{ a.s.,}
		\end{multline*}
		by definition of essential supremum. Besides, we have by Lemma~\ref{lem:FatouL} that, for every $k\in\mathbb{N}$, %
		\begin{equation*}
			V^{+}\!\left(H_{n_{k}\!\left(\cdot\right)}\!\left(\cdot\right)+\left\langle \zeta_{n_{k}\!\left(\cdot\right)}\!\left(\cdot\right),Y\!\left(\cdot\right) \right\rangle,\cdot\right)\leq V^{+}\!\left(M+\left\langle \zeta_{n_{k}\!\left(\cdot\right)}\!\left(\cdot\right),Y\!\left(\cdot\right) \right\rangle,\cdot\right)\leq L_{M}\!\left(\cdot\right) \text{ a.s.} %
		\end{equation*}
		(note that $\zeta_{n_{k}}\in\Xi^{d}\!\left(M\right)$), so the limsup Fatou lemma yields (cf.\ Assumption~\ref{as:esssupfiniteas})
		\begin{align*}
			\lefteqn{\mathbb{E}_{\mathbb{P}}\!\left[\left.V\!\left(H\!\left(\cdot\right)+\left\langle \zeta\!\left(\cdot\right),Y\!\left(\cdot\right) \right\rangle,\cdot\right)\right|\mathscr{G}\right]}\\
			&\hspace{3cm}\geq \limsup_{k\rightarrow+\infty}\mathbb{E}_{\mathbb{P}}\!\left[\left.V\!\left(H_{n_{k}\!\left(\cdot\right)}\!\left(\cdot\right)+\left\langle \zeta_{n_{k}\!\left(\cdot\right)}\!\left(\cdot\right),Y\!\left(\cdot\right) \right\rangle,\cdot\right)\right|\mathscr{G}\right]\\
			&\hspace{3cm}\geq \limsup_{k\rightarrow+\infty}\left(G\!\left(H_{n_{k}\!\left(\cdot\right)}\!\left(\cdot\right),\cdot\right) -\frac{1}{n_{k}\!\left(\cdot\right)} \right)=G\!\left(H\!\left(\cdot\right),\cdot\right)\text{ a.s.,}
		\end{align*}
		where the last inequality follows from \eqref{neige}. Combining the inequalities above, we establish \eqref{eq:GH} for any bounded $H$ as well.
			
		Finally, we extend the above result to an arbitrary $\mathscr{G}$-measurable $H\geq 0$ a.s. Since $H=\sum_{n\in\mathbb{N}}H_{n}$, with each $H_{n}\triangleq H\bbOne_{\left\{n-1\leq H<n\right\}}$ $\mathscr{G}$-measurable and bounded, %we have
%		\begin{align*}
%			 G\!\left(H\!\left(\cdot\right),\cdot\right)&=\sum_{n\in\mathbb{N}}G\!\left(H_{n}\!\left(\cdot\right),\cdot\right)\bbOne_{\left\{n-1\leq H<n\right\}}\\
%			 &=\sum_{n\in\mathbb{N}}\esssup_{\xi\in\Xi^{d}\!\left(H_{n}\right)}\mathbb{E}_{\mathbb{P}}\!\left[\left.V\!\left(H_{n}\!\left(\cdot\right)+\left\langle \xi\!\left(\cdot\right),Y\!\left(\cdot\right) \right\rangle,\cdot\right)\right|\mathscr{G}\right]\bbOne_{\left\{n-1\leq H<n\right\}}\text{ a.s.}
%		\end{align*}
		we can obtain the desired equality from the bounded case.
		
		\item
		Lastly, we claim that almost all paths of $G$ are left-continuous. This will be done by constructing some left-continuous function $\overline{G}$ such that 
$$
\mathbb{P}\!\left\{\omega\in {\Omega}\text{: }\forall\,x\geq 0,\ G\!\left(x,\omega\right)=\overline{G}\!\left(x,\omega\right)\right\}=1.
$$
%To see this, let us
%first remark that the function $G\!\left(x,\cdot\right):~\Omega\rightarrow \mathbb{R}$ is
%$\mathscr{G}$-measurable as already mentioned right after its definition. %begin with the remark that, as shown above, for every $x\geq 0$, the function $G\!\left(x,\cdot\right):~\Omega\rightarrow \mathbb{R}$, being a version of the essential supremum of $\mathscr{G}$-measurable random variables, is itself measurable with respect to $\mathscr{G}$.
%	In addition, a.e.\ path of $G$ is right-continuous. Therefore, for every $\omega$ outside a $\mathbb{P}$-zero set, we have that $\lim_{n\rightarrow+\infty}G_{n}\!\left(x,\omega\right)=G\!\left(x,\omega\right)$ for all $x\geq0$, with
%	\begin{equation*}
%		G_{n}\!\left(x,\omega\right)\triangleq\sum_{k\in\mathbb{Z}} G\!\left(k/2^n,\omega\right)\bbOne_{\left[\left.k/2^n,\left(k+1\right)/2^n\right)\right.\times\Omega}\!\left(x,\omega\right).
%	\end{equation*}
%	Hence $G:~\left[\left.0,+\infty\right)\right.\times\Omega\rightarrow \mathbb{R}$ is measurable with
%respect to the product $\sigma$-algebra $\mathscr{B}\!\left(\left[\left.0,+\infty\right)\right.\right)\otimes\mathscr{G}$ since
%each of the $G_{n}$ clearly are.
	
	We define
	\begin{equation*}
		\overline{G}\!\left(x,\omega\right)\triangleq \left\{
			\begin{array}{ll}
				\sup_{\substack{q\in\mathbb{Q}\\q<x}} G\!\left(q,\omega\right),& \text{if }x>0,\\
				G\!\left(0,\omega\right),& \text{otherwise.}
			\end{array}
		\right.
	\end{equation*}
From \emph{(iii)} recall that $G$ is $\mathscr{B}\!\left(\left[\left.0,+\infty\right)\right.\right)\otimes\mathscr{G}$-measurable, so 	it is obvious that $\overline{G}$ %:~\left[\left.0,+\infty\right)\right.\times\Omega\rightarrow \mathbb{R}\cup\left\{-\infty,+\infty\right\}$ %
	is $\mathscr{B}\!\left(\left[\left.0,+\infty\right)\right.\right)\otimes\mathscr{G}$-measurable too. Besides, it is trivial to check that, for every $\omega\in\Omega$, the function $\overline{G}\!\left(\cdot,\omega\right)$ %:~\left[\left.0,+\infty\right)\right.\rightarrow \mathbb{R}\cup\left\{-\infty,+\infty\right\}$ %
		is non-decreasing on $\left(0,+\infty\right)$. We remark further that, by construction, all paths of $\overline{G}$ are left-continuous on $\left(0,+\infty\right)$.
		
		It follows immediately from the monotonicity of all the sample paths of $G$ that the inequality $G\!\left(x,\omega\right)\geq\overline{G}\!\left(x,\omega\right)$ holds true simultaneously for every $x\geq 0$, for every $\omega\in\Omega$. In particular, this gives that, for every $\omega\in\Omega$ and for all $x\geq0$, it holds that $\overline{G}\!\left(x,\omega\right)<+\infty$. At last, we shall show that $\mathbb{P}\!\left\{\omega\in {\Omega}\text{: }\forall\,x\geq 0,\ G\!\left(x,\omega\right)=\overline{G}\!\left(x,\omega\right)\right\}=1$.
		\begin{enumerate}[label=\emph{(\alph*)}]
			\item
			The proof is by contradiction. Let us suppose that the set
			\begin{equation*}
				\Omega_{1}\triangleq \left\{\omega\in {\Omega}\text{ : }\exists\,x>0\text{ s.t.\ }G\!\left(x,\omega\right)>\overline{G}\!\left(x,\omega\right)\right\}
			\end{equation*}
			has strictly positive measure, i.e., $\mathbb{P}\!\left(\Omega_{1}\right)>0$. Note that, because $\left(\Omega,\mathscr{G},\mathbb{P}\right)$ is a complete measure space, we can apply the measurable projection theorem (see
e.g.\ Theorem~4 in \citet{stbeuve})
%e.g.\ Theorem~III.23 in \citet{castaingvaladier77})
to deduce that%
			\footnote{
				Given a set $E\subseteq X\times Y$, we recall that the \emph{projection of $E$ on $X$} is
				\begin{equation*}
					\Proj[X]{E}\triangleq\left\{x\in X\text{: }\exists\,y\in Y\text{ such that }\left(x,y\right)\in E\right\}.
				\end{equation*}}
			%\begin{align*}
			%	\Omega_{1}&=\left\{\omega\in\Omega\text{: }\exists\,x\geq 0 \text{ s.t.\ }\left(\omega,x\right)\in\left(G-\overline{G}\right)^{-1}\!\left(\left(0,+\infty\right)\right)\right\}\\
			%	&=\Proj{\left(G-\overline{G}\right)^{-1}\!\left(\left(0,+\infty\right)\right)}
			%\end{align*}
			$\Omega_{1}=\Proj{\left(G-\overline{G}\right)^{-1}\!\left(\left(0,+\infty\right)\right)}$ %
			belongs to $\mathscr{G}$.
			
			Consider the multi-function $\mathscr{E}:~\Omega\rightrightarrows\left[\left.0,+\infty\right)\right.$ given by
			\begin{equation*}
				\mathscr{E}\!\left({\omega}\right)\triangleq \left\{
					\begin{array}{ll}
						\left\{x>0\text{: }G\!\left(x,\omega\right)>\overline{G}\!\left(x,\omega\right)\right\},& \text{if } \omega\in\Omega_{1},\\
						$\{1\}$,& \text{otherwise.}
					\end{array}
				\right.
			\end{equation*}
Its graph is given by
			\begin{equation*}
				 \gph\mathscr{E}=\left(\Omega_{1}^{c}\times\left\{1\right\}\right)\cup\left(\left[\Omega_{1}\times\left(0,+\infty\right)\right]\cap\left[\left(G-\overline{G}\right)^{-1}\!\left(\left(0,+\infty\right)\right)\right]\right)
			\end{equation*}
and belongs to $\mathscr{B}\!\left(\left[\left.0,+\infty\right)\right.\right)\otimes\mathscr{G}$.
%is a $\mathscr{G}$-random set.
Consequently, we can apply the von Neumann-Aumann theorem (see
e.g.\ Theorem~3 in \citet{stbeuve})
%\Cref{th:NeumannAumann} %
			to produce a $\mathscr{G}$-measurable selector $H:~{\Omega}\rightarrow\left[\left.0,+\infty\right)\right.$ of $\mathscr{E}$. %
			%on $\{\omega| \mathscr{E}(\omega) \neq \emptyset \}$.
			In particular, this implies that
			\begin{equation}\label{eq:G>Gbar}
				\mathbb{P}\left\{\omega\in {\Omega}\text{: }G\!\left(H\!\left(\omega\right),\omega\right)>\overline{G}\!\left(H\!\left(\omega\right),\omega\right)\right\}\geq\mathbb{P}\!\left(\Omega_{1}\right)>0.
			\end{equation}
			
As $G\!\left(0,\omega\right)=\overline{G}\!\left(0,\omega\right)$, we get that $H>0$. Furthermore, we may and shall assume, without loss of generality,
			that there exists some $\varepsilon\in\left.\left(0,1\right.\right]$ such that
			$H>\varepsilon$.
			
			\item
			On the other hand, we shall see that $G\!\left(H\!\left(\omega\right),\omega\right)\leq \overline{G}\!\left(H\!\left(\omega\right),\omega\right)$ holds for $\mathbb{P}$-a.e.\ $\omega\in\Omega$, contradicting \eqref{eq:G>Gbar}.
			
			Firstly, fix an arbitrary $n\in\mathbb{N}$. As in part \ref{partversion} of this proof, it is possible to construct some $\zeta_{n}\in\Xi^{d}\!\left(H\right)$ such that, for $\mathbb{P}$-a.e.\ $\omega\in\Omega$,
			\begin{equation}
\label{brise}
				\mathbb{E}_{\mathbb{P}}\!\left[\left.V\!\left(H\!\left(\cdot\right)+\left\langle \zeta_{n}\!\left(\cdot\right),Y\!\left(\cdot\right) \right\rangle,\cdot\right)\right|\mathscr{G}\right]\!\left(\omega\right)\geq G\!\left(H\!\left(\omega\right),\omega\right)-\frac{1}{n}.
			\end{equation}
			%holds for $\mathbb{P}$-a.e.\ $\omega\in\Omega$.
			
			Next, setting for every $m\in\mathbb{N}$ (recall that $H>\varepsilon$),
			\begin{equation*}
				f_{n}^{m}\!\left(\omega\right)\triangleq V\!\left(H\!\left(\omega\right)-\frac{\varepsilon}{m}+\frac{H\!\left(\omega\right)-\varepsilon/m}{H\!\left(\omega\right)}\left\langle \zeta_{n}\!\left(\omega\right),Y\!\left(\omega\right) \right\rangle,\omega\right),\qquad \omega\in\Omega,
			\end{equation*}
			it is trivial by continuity (see Assumption~\ref{as:V}) that %the sequence %of random variables %
			$\left\{f_{n}^{m}\text{; }m\in\mathbb{N}\right\}$ converges a.s.\ to $V\!\left(H\!\left(\cdot\right)+\left\langle \zeta_{n}\!\left(\cdot\right),Y\!\left(\cdot\right) \right\rangle,\cdot\right)$ as $m\rightarrow+\infty$. Thus, Fatou's lemma gives
			\begin{equation*}
				 \liminf_{m\rightarrow+\infty}\mathbb{E}_{\mathbb{P}}\!\left[\left.\left[f_{n}^{m}\right]^{+}\right|\mathscr{G}\right]\geq \mathbb{E}_{\mathbb{P}}\!\left[\left.V^{+}\!\left(H\!\left(\cdot\right)+\left\langle \zeta_{n}\!\left(\cdot\right),Y\!\left(\cdot\right) \right\rangle,\cdot\right)\right|\mathscr{G}\right]\text{ a.s.}
			\end{equation*}
			
			Secondly, we note that, for each $m\in\mathbb{N}$, the random vector $\zeta_{n}\left(H-\varepsilon/m\right)/H$ belongs to $\Xi^{d}\!\left(H-\varepsilon/m\right)$, because
			\begin{equation*}
				H-\frac{\varepsilon}{m}+\left\langle \frac{H-\varepsilon/m}{H}\zeta_{n},Y \right\rangle=\frac{H-\varepsilon/m}{H}\left(H+\left\langle \zeta_{n},Y \right\rangle\right)\geq 0\text{ a.s.}
			\end{equation*}
			(recall that $H>\varepsilon$ and $\zeta_{n}\in\Xi^{d}\!\left(H\right)$).
			
			Therefore, given Assumption~\ref{as:V-} and the fact that, for every $m\in\mathbb{N}$,
			the inequality $\left[f_{n}^{m}\right]^{-}\leq V^{-}\!\left(0,\cdot\right)$ is true a.s.,
			we can apply the limsup Fatou lemma to obtain
			\begin{equation*}
			\limsup_{m\rightarrow+\infty}\mathbb{E}_{\mathbb{P}}\!\left[\left.\left[f_{n}^{m}\right]^{-}\right|\mathscr{G}\right]\leq \mathbb{E}_{\mathbb{P}}\!\left[\left.V^{-}\!\left(H\!\left(\cdot\right)+\left\langle \zeta_{n}\!\left(\cdot\right),Y\!\left(\cdot\right) \right\rangle,\cdot\right)\right|\mathscr{G}\right]\text{ a.s.}
			\end{equation*}
			Combining both inequalities yields
			\begin{equation*}
				\liminf_{m\rightarrow+\infty}\mathbb{E}_{\mathbb{P}}\!\left[\left.f_{n}^{m}\right|\mathscr{G}\right]\geq \mathbb{E}_{\mathbb{P}}\!\left[\left.V\!\left(H\!\left(\cdot\right)+\left\langle \zeta_{n}\!\left(\cdot\right),Y\!\left(\cdot\right) \right\rangle,\cdot\right)\right|\mathscr{G}\right] \text{ a.s.,}
			\end{equation*}
			and from \eqref{brise} we get
			\begin{equation}\label{bruine}
				\liminf_{m\rightarrow+\infty}\mathbb{E}_{\mathbb{P}}\!\left[\left.f_{n}^{m}\right|\mathscr{G}\right] \geq G\!\left(H\!\left(\cdot\right),\cdot\right)-\frac{1}{n}\text{ a.s.}
			\end{equation}
			Besides, for every $m\in\mathbb{N}$ we have that
			\begin{equation*}
				 \esssup_{\xi\in\Xi^{d}\!\left(H-\varepsilon/m\right)}\mathbb{E}_{\mathbb{P}}\!\left[\left.V\!\left(H\!\left(\cdot\right)-\frac{\varepsilon}{m}+\left\langle \xi\!\left(\cdot\right),Y\!\left(\cdot\right) \right\rangle,\cdot\right)\right|\mathscr{G}\right]\geq\mathbb{E}_{\mathbb{P}}\!\left[\left.f_{n}^{m}\right|\mathscr{G}\right]\text{ a.s.,}
			\end{equation*}
			and so
			\begin{multline*}
				 \liminf_{m\rightarrow+\infty}\esssup_{\xi\in\Xi^{d}\!\left(H-\varepsilon/m\right)}\mathbb{E}_{\mathbb{P}}\!\left[\left.V\!\left(H\!\left(\cdot\right)-\frac{\varepsilon}{m}+\left\langle \xi\!\left(\cdot\right),Y\!\left(\cdot\right) \right\rangle,\cdot\right)\right|\mathscr{G}\right]\\
				\geq\liminf_{m\rightarrow+\infty}\mathbb{E}_{\mathbb{P}}\!\left[\left.f_{n}^{m}\right|\mathscr{G}\right]\text{ a.s.}
			\end{multline*}
			
			On the other hand, let $m\in\mathbb{N}$ be arbitrary, but fixed. Then we know by the preceding step that
			\begin{equation*}
				 \esssup_{\xi\in\Xi^{d}\!\left(H-\varepsilon/m\right)}\mathbb{E}_{\mathbb{P}}\!\left[\left.V\!\left(H\!\left(\cdot\right)-\frac{\varepsilon}{m}+\left\langle \xi\!\left(\cdot\right),Y\!\left(\cdot\right) \right\rangle,\cdot\right)\right|\mathscr{G}\right]\!\left(\omega\right)=G\!\left(H\!\left(\omega\right)-\frac{\varepsilon}{m},\omega\right)
			\end{equation*}
			for every $\omega$ outside a $\mathbb{P}$-null set. Using \eqref{bruine}, we get
			\begin{equation}\label{soleil}
				\liminf_{m\rightarrow+\infty}G\!\left(H\!\left(\cdot\right)-\frac{\varepsilon}{m},\cdot\right) \geq G\!\left(H\!\left(\cdot\right),\cdot\right)-\frac{1}{n}\text{ a.s.}
			\end{equation}
			Next, choosing $q_{m}(\omega)\in\mathbb{Q}$, $q_m(\omega)>0$ such that $H\!\left(\omega\right)-\varepsilon/m\leq q_{m}(\omega)<H\!\left(\omega\right)$, it follows immediately from the definition of $\overline{G}$ (recall that $H>\varepsilon>0$) and from the monotonicity of $G$ (see the first part of this proof) that
			\begin{align*}
				\overline{G}\!\left(H\!\left(\omega\right),\omega\right)=\sup_{\substack{q\in\mathbb{Q}\\q<H\!\left(\omega\right)}} G\!\left(q,\omega\right)&\geq G\!\left(q_{m}(\omega),\omega\right)\\
				&\geq G\!\left(H\!\left(\omega\right)-\varepsilon/m,\omega\right)\geq \inf_{k\geq m}G\!\left(H\!\left(\omega\right)-\varepsilon/k,\omega\right),
			\end{align*}
			consequently,
			\begin{equation*}
				\overline{G}\!\left(H\!\left(\cdot\right),\cdot\right)\geq\sup_{m\in\mathbb{N}}\inf_{k\geq m}G\!\left(H\!\left(\cdot\right)-\varepsilon/k,\cdot\right)=\liminf_{m\rightarrow+\infty}G\!\left(H\!\left(\cdot\right)-\varepsilon/m,\cdot\right)\text{ a.s.}
			\end{equation*}			
			
			So, from \eqref{soleil}, for every $n\in\mathbb{N}$, %
			%\begin{equation*}
			$	\overline{G}\!\left(H\!\left(\cdot\right),\cdot\right)\geq G\!\left(H\!\left(\cdot\right),\cdot\right)-1/n$ a.s., %\text{ a.s.},
			%\end{equation*}
			hence
			\begin{equation*}
				\overline{G}\!\left(H\!\left(\cdot\right),\cdot\right)\geq \limsup_{n\rightarrow+\infty}\left(G\!\left(H\!\left(\cdot\right),\cdot\right)-\frac{1}{n}\right)=G\!\left(H\!\left(\cdot\right),\cdot\right)\text{ a.s.},
			\end{equation*}
			as claimed.\qedhere
		\end{enumerate}
	\end{enumerate}
\end{proof}

\begin{proof}[Proof of Proposition~\ref{prop:optonestep}]
	As in \ref{partversion} of the previous proof, one can show that there exists some sequence $\xi_{n}\!\left(\cdot\right)\in\Xi^{d}\!\left(H\right)$ that attains the essential supremum in \eqref{eq:GH}. We may assume that $\xi_{n}\!\left(\cdot\right)\in D(\cdot)$ a.s., and hence for every $n\in\mathbb{N}$ we have by Lemma~\ref{lem:rvKx} that $\left\Vert\xi_{n}\!\left(\cdot\right)\right\Vert\leq H\!\left(\cdot\right)/\beta\!\left(\cdot\right)$ a.s. Next, Lemma~2 of \citet{kabanov01} implies the existence of a $\mathscr{G}$-measurable random subsequence $\left\{\xi_{n_{k}}\text{; }k\in\mathbb{N}\right\}$ such that $\lim_{k\rightarrow+\infty}\xi_{n_{k}\!\left(\cdot\right)}\!\left(\cdot\right)=\xi\!\left(\cdot\right)$ a.s.\ for some $\xi\!\left(\cdot\right)\in\Xi^{d}\!\left(H\right)$. Lemmata~\ref{lem:FatouL} and \ref{lem:versionesssup} allow the use of the (conditional) Fatou lemma, hence we get that $\widetilde{\xi}\!\left(H\right)\left(\cdot\right)\triangleq\xi\!\left(\cdot\right)$ is as claimed.
\end{proof}

%=========================================================================================================================================

%%%%%%%%%%%%%%%%%%%%%%%%%%%%%%%%%%%%%%%%%%%%%%%%%%%%%%%%%%%%%%%%%%%
%%                                                               %%
%% References																										 %%
%%                                                               %%
%%%%%%%%%%%%%%%%%%%%%%%%%%%%%%%%%%%%%%%%%%%%%%%%%%%%%%%%%%%%%%%%%%%

\end{document}